\documentclass[preprint2,twocolappendix]{aastex63}
\usepackage{CJK}
\usepackage{float}
\usepackage[absolute]{textpos}
\usepackage{amsmath}
\usepackage{mathrsfs}
\usepackage{txfonts}
\usepackage{bm}

\shorttitle{SCCC in L1251}
\shortauthors{Liu \& Wu et al.}

\begin{document}
\begin{CJK}{UTF8}{gbsn}

\title{A search for cloud cores affected by shocked carbon chain chemistry in L1251}

\email{
ywu@pku.edu.cn\\
1501110219@pku.edu.cn}

\author{X.-C. Liu (刘训川)}
\affiliation{Department of Astronomy, School of Physics, Peking University, 100871 Beijing, China}
\affiliation{Kavli Institute for Astronomy and Astrophysics, Peking University, 100871 Beijing, China}

\author{Y. Wu}
\affiliation{Department of Astronomy, School of Physics, Peking University, 100871 Beijing, China}
\affiliation{Kavli Institute for Astronomy and Astrophysics, Peking University, 100871 Beijing, China}

\author{C. Zhang (张超)}
\affiliation{Department of Astronomy, School of Physics, Peking University, 100871 Beijing, China}
\affiliation{Department of Astronomy, Yunnan University, Kunming, 650091, China}

\author{X. Chen}
\affiliation{Center for Astrophysics, GuangZhou University, Guangzhou 510006, China}

\author{L.-H. Lin}
\affiliation{{Purple Mountain Observatory, Chinese Academy of Sciences, 10 Yuanhua Road, Nanjing 210033, P.R. China}}
\affiliation{{Key Laboratory of Radio Astronomy, Chinese Academy of Sciences, 10 Yuanhua Road, Nanjing 210033, P.R. China}}
\affiliation{{School of Astronomy and Space Science, University of Science and Technology of China, 96 Jinzhai Road, Hefei 230026, P.R. China}}

\author{S.-L. Qin}
\affiliation{Department of Astronomy, Yunnan University, Kunming, 650091, China}

\author{T. Liu}
\affiliation{Shanghai Astronomical Observatory, Chinese Academy of Sciences, Shanghai 200030, China}

\author{C. Henkel}
\affiliation{Max-Planck Institut f\"{u}r Radioastronomie, Auf dem H\"{u}gel 69, 53121 Bonn, Germany}
\affiliation{Astron. Dept., King Abdulaziz University, P.O. Box 80203, Jeddah 21589, Saudi Arabia}

\author{J. Wang}
\affiliation{Shanghai Astronomical Observatory, Chinese Academy of Sciences, Shanghai 200030, China}

\author{H.-L. Liu}
\affiliation{Departamento de Astronom\'ia, Universidad de Concepci\'on, Av. Esteban Iturra s/n, Distrito Universitario, 160-C, Chile}

\author{J. Yuan}
\affiliation{National Astronomical Observatories, Chinese Academy of Sciences, Beijing 100101, China}

\author{L.-X. Yuan}
\affiliation{National Astronomical Observatories, Chinese Academy of Sciences, Beijing 100101, China}

\author{J. Li}
\affiliation{Shanghai Astronomical Observatory, Chinese Academy of Sciences, Shanghai 200030, China}

\author{Z.-Q. Shen}
\affiliation{Shanghai Astronomical Observatory, Chinese Academy of Sciences, Shanghai 200030, China}

\author{D. Li}
\affiliation{National Astronomical Observatories, Chinese Academy of Sciences, Beijing 100101, China}
\affiliation{NAOC-UKZN Computational Astrophysics Centre, University of KwaZulu-Natal, Durban 4000, South Africa}

\author{J. Esimbek}
\affiliation{Xinjiang Astronomical Observatory, Chinese Academy of Sciences, 830011, Urumqi, China}

\author{K. Wang}
\affiliation{Kavli Institute for Astronomy and Astrophysics, Peking University, 100871 Beijing, China}
\affiliation{Department of Astronomy, School of Physics, Peking University, 100871 Beijing, China}

\author{L.-X. Li}
\affiliation{Kavli Institute for Astronomy and Astrophysics, Peking University, 100871 Beijing, China}
\affiliation{Department of Astronomy, School of Physics, Peking University, 100871 Beijing, China}

\author{Kee-Tae Kim}
\affiliation{Korea Astronomy and Space Science Institute, 776 Daedeokdae-ro, Yuseong-gu, Daejeon 34055, Republic of Korea}
\affiliation{University of Science and Technology, Korea (UST), 217 Gajeong-ro, Yuseong-gu, Daejeon 34113, Republic of Korea}

\author{L. Zhu}
\affiliation{National Astronomical Observatories, Chinese Academy of Sciences, Beijing 100101, China}

\author{D. Madones}
\affiliation{Department of Astronomy, University of Chile, Casilla 36-D, Santiago, Chile}

\author{N. Inostroza}
\affiliation{N\'{u}cleo de Astroqu\'{i}micay Astrof\'{i}sica, Instituto de Ciencias Qu\'{i}micas Aplicadas, Facultad de Ingenier\'{i}a, Universidad Aut\'{o}noma de Chile Av.}

\author{F.-Y. Meng}
\affiliation{I. Physikalisches Institut, Universit\"{a}t z\"{u} K\"{o}ln, Z\"{u}lpicher Str. 77, D-50937 K\"{o}ln, Germany}

\author{T. Zhang}
\affiliation{I. Physikalisches Institut, Universit\"{a}t z\"{u} K\"{o}ln, Z\"{u}lpicher Str. 77, D-50937 K\"{o}ln, Germany}

\author{K. Tatematsu}
\affiliation{Nobeyama Radio Observatory, National Astronomical Observatory of Japan, National Institutes of Natural Sciences, 462-2 Nobeyama,
   Minamimaki, Minamisaku, Nagano 384-1305, Japan}

\author{Y. Xu}
\affiliation{Purple Mountain Observatory, Qinghai Station, 817000, Delingha, China}

\author{B.-G. Ju}
\affiliation{Purple Mountain Observatory, Qinghai Station, 817000, Delingha, China}

\author{A. Kraus}
\affiliation{NAOC-UKZN Computational Astrophysics Centre, University of KwaZulu-Natal, Durban 4000, South Africa}

\author{F.-W. Xu}
\affiliation{Department of Astronomy, School of Physics, Peking University, 100871 Beijing, China}
\affiliation{Kavli Institute for Astronomy and Astrophysics, Peking University, 100871 Beijing, China}

\begin{abstract}
We searched for shocked carbon chain chemistry (SCCC) sources with C$_3$S abundances surpassing those of 
HC$_5$N towards the dark cloud L1251, using the Effelsberg telescope at K-band (18 -- 26\,GHz). 
L1251-1 and L1251-3 are identified as the most promising SCCC sources. The two sources harbor young stellar objects.
We conducted mapping observations towards L1251-A, 
the western tail of L1251, at $\lambda$ $\sim$3\,mm with the PMO 13.7 m and the NRO 45 m 
telescopes in lines of C$_2$H, N$_2$H$^+$, CS, HCO$^+$, SO, HC$_3$N and C$^{18}$O as well as in CO 3--2 using 
the JCMT. The spectral data were combined with archival data including Spitzer and Herschel continuum maps
for further analysis.
Filamentary sub-structures labeled as F1 to F6 were extracted in L1251, with F1 being  associated with 
L1251-A hosting L1251-1. The peak positions of dense gas traced by HCO$^+$ are misaligned relative to those 
of the dust clumps.  Episodic outflows are common in this region. The twisted morphology of F1 and velocity 
distribution along L1251-A may originate from stellar feedback. SCCC in L1251-1 may have been caused 
by outflow activities originated from the infrared source IRS1. 
The signposts of ongoing
SCCC and the broadened line widths of C$_3$S and C$_4$H in L1251-1 as well as the distribution of HC$_3$N 
are also related to outflow activities in this region.
L1251-1 (IRS1) together with the previously identified
SCCC source IRS3 demonstrate that L1251-A is an excellent region to study shocked carbon chain 
chemistry.
\end{abstract}



\section{Introduction}
Carbon chain molecules (CCMs) are particularly interesting since they are important tracers of organic 
interstellar chemistry. The nitrogen-bearing CCMs  HC$_3$N $J$ = 1--0 and HC$_5$N $J$ = 4--3 were 
firstly detected towards Sgr B2 in the  seventies \citep{1971ApJ...163L..35T,1976ApJ...205...82M}. Soon 
after, HC$_7$N and HC$_9$N were found in TMC-1 \citep{1978ApJ...219L.133K,1978ApJ...223L.105B}. They are 
important players in hydrocarbon chemistry \citep{2001ApJ...558..693D} and constrain evolutionary stages 
of dark cold cores \citep{2012MNRAS.419..238B}. These molecular species can also trace dynamical motions 
in molecular clouds, including material infall \citep{2013MNRAS.436.1513F}. In some cold 
clouds cyanopolyynes (HC$_{2n+1}$N) are particularly prominent and even HC$_{11}$N has been detected in TMC-1. 
\citep{2020arXiv200911900L}. 
In star forming regions, cyanopolyynes  are usually suppressed, although they 
can be abundant in some massive star forming regions \citep{2019ApJ...881...57T}. 

Another important kind of 
unsaturated CCMs, sulfur-bearing molecules including C$_2$S and C$_3$S, 
can be used to investigate physical conditions of their production 
regions. In early cold cores, the column densities of sulfur-bearing molecules are usually positively correlated with those of 
nitrogen-containing molecules \citep{1992ApJ...392..551S,1992ApJ...394..539H,2016A&A...593A..94F}. 
In embedded low-mass protostars, significant positive correlations between the cyanopolyynes (HC$_\mathrm{n}$N) and pure 
hydrocarbon chains (C$_\mathrm{n}$H) were found \citep{2018ApJ...863...88L}.
C2S lines were not or only marginally detected in a number of low mass star formation cores  \citep{1992ApJ...392..551S}.

In L1527 (IRAS\,04368+2557), \citet{2008ApJ...675L..89S,2008ApJ...672..371S} found strong emissions 
of high-extinction lines ($E_{\rm u}>$ 20 K) of unsaturated hydrocarbons, including C$_4$H$_2$, C$_4$H and  $l\text{-}$C$_3$H$_2$. A mechanism named 
warm carbon chain chemistry (WCCC) was put forward to explain the phenomena. In the picture of WCCC, 
prestellar objects contract and heat the surrounding dust, and the evaporated CH$_4$ reacts with C$^+$ to 
produce CCMs \citep{2008ApJ...672..371S,2009ApJ...697..769S}. The second WCCC source (I15398-3359) was 
found in Lupus I \citep{2009ApJ...697..769S}. Molecular outflows were found in both WCCC sources 
\citep{2008ApJ...672..371S,2014ApJ...795..152O,2015A&A...576A.109Y}. The excitation temperatures 
of hydrocarbons such as C$_4$H$_2$ and CH$_3$CCH  are determined to be about 
12 K in both WCCC sources \citep{2008ApJ...675L..89S,2009ApJ...697..769S}, which 
are lower than the sublimation temperature of CH$_4$ (25 K) but much higher than the excitation temperature 
of hydrocarbons in TMC-1 \citep[3.8--6.7 K for C$_4$H$_2$;][]{2004PASJ...56...69K,2008ApJ...672..371S}.

Recently we have observed six outflows and five starless Lupus cores using the Shanghai 65 m Tian Ma Radio 
Telescope (TMRT) to explore the production region and excitation mechanism of CCMs \citep{2019MNRAS.488..495W}. 
All the targets are located within 330 pc from the Earth. The HC$_3$N $J$ = 2--1, HC$_5$N $J$ = 6--5, HC$_7$N $J$ = 14--13, 
15--14, 16--15 and C$_3$S $J$ = 3--2 transitions were observed.  
These lines were detected towards ten of the eleven observed sources.
Three sources, IRAS 20582+7724 (I20582), L1221 and L1251A (corresponding to IRS3 in the L1251-A region,
hereafter L1251-IRS3), 
were explained in terms of a new chemical 
mechanism -- shocked carbon chain chemistry (SCCC); another outflow source, the eastern molecular core of the outflow
source L1489 (L1489 EMC) was identified as a particular carbon-chain production region and as a candidate 
of a WCCC source \citep{2019A&A...627A.162W}. 

The main concept of SCCC is that the sulfur ions and atoms
released from grain surfaces under the effect of shocks will mix with cold gas components and drive a new 
generation of carbon chain molecules, especially the sulfur-bearing ones such as C$_3$S. The criteria of SCCC 
include relatively weak N-bearing CCM emissions in contrast to strong C$_3$S emissions and signs of 
developed shock regions. Our model \citep{2019MNRAS.488..495W}  forecasts that, in sources where the 
SCCC mechanism works most efficiently, the abundance of C$_3$S (X(C$_3$S)) will not only exceed that of 
HC$_7$N (X(HC$_7$N)), but also that of HC$_5$N (X(HC$_5$N)). The observations presented by 
\citet{2019MNRAS.488..495W} show that  so far known three SCCC sources (IRAS 20582+7724 (I20582), L1221 and 
L1251-IRS3) all have $N$(C$_3$S)$<$$N$(HC$_5$N), while L1251-IRS3 
possesses a C$_3$S column density closest to but still not exceeding that of HC$_5$N.

In this work, we further search for and  examine SCCC sources in the L1251 region (see Table \ref{tab:4stars}), using the 
Effelsberg 100 m telescope in K-band (18--26 GHz). We mapped multiple lines towards  the most 
promising SCCC candidate (L1251-1) located in the tail of the L1251 region, using the Purple Mountain Observatory  
(PMO) 13.7 m and the Nobeyama Radio Observatory (NRO) 45 m telescopes in the 3 mm band as well as the James 
Clerk Maxwell Telescope (JCMT) in CO $J$ = 3--2 (see Sect. \ref{sec:obs}). In Sect. \ref{sec_L1251}, we introduce L1251, 
including its 
morphology and structure as well as the star formation activities in L1251-A. The results of our search for 
new SCCC sources in the L1251 region are presented in Sect. \ref{sec:singleobs}, while results of 
further mapping observations towards L1251-A are presented in Sect. \ref{sec:mapping}. In Sect. \ref{sec:dis}, 
we discuss filamentary structures in L1251-A and SCCC related phenomena in this region. 
Sect. \ref{sec:summary} provides a summary.

\section{Observations} \label{sec:obs}
\subsection{Effelsberg}
Single point observations of L1251-1, L1251-2 and L1251-3  were performed during 2018 
Jan 29 to 2018 Feb 2 using the Effelsberg 100 m telescope  
(see Table \ref{tab:4stars} and Fig. \ref{herscmap} for their 
coordinates and locations) in the L1251 region. A K-band (18--26 GHz) HEMT receiver with two polarizations 
(LCP/RCP) was used as the frontend. The lines covered by the K-band are shown in Table \ref{table_lines_teles}. 
Integration times ranged from 40 to 80 minutes, depending on the system temperature and the strength of the 
emission lines. The main beam efficiency ranges from 0.61 to 0.79 depending on the 
frequency\footnote{\url{https://eff100mwiki.mpifr-bonn.mpg.de/doku.php?id=information_for_astronomers:rx:s14mm}}.
Flux calibration was accurate to 10\%, estimated by observing the standard source NGC7027 
\citep{1994A&A...284..331O}. The pointing accuracy was better than 5\arcsec. The half power beam width (HPBW) of 
the telescope at the observed frequency was approximately 40\arcsec (0.06 pc assuming a distance of 300 pc). 
Four subbands, WFF4 (18--20.5 GHz), WFF3 
(20--22.5 GHz), WFF2(21.6--24.1) and WFF1(23.5--26), can simultaneously cover the whole K-band (18--26 GHz). All 
spectra were smoothed to a channel spacing of about 0.5 km s$^{-1}$. The 1-$\sigma$ rms noise varied from one 
to several tens of mK, mainly because of different  on-source integration times for our targets.

For unknown reasons, the spectra of WFF3 are shifted redwards by approximately 1 km s$^{-1}$. These misalignments are 
systematic for all the observed sources and both the two lines c-C$_3$H$_2$ 2$_{2,0}$--2$_{1,1}$ and HC$_5$N $J$ = 
8--7 located within WFF3. Therefore, the WFF3 spectra measured were shifted bluewards by two channels.

\subsection{PMO 13.7 m}
C$_2$H $N$ = 1--0, N$_2$H$^+$ $J$ = 1--0 and CS $J$ = 2--1 were mapped towards L1251-A (see Fig. \ref{herscmap})
in the On-The-Fly (OTF) mode. The maps cover an area of 20\arcmin$\times$10\arcmin~and using the PMO 13.7 m telescope.
The PMO 13.7 m telescope employs a nine-beam Superconducting Spectroscopic Array Receiver (SSAR) as the front 
end in sideband separation mode \citep[see][]{2012ITTST...2..593S}. A Fast Fourier Transform Spectrometer (FFTS) 
was employed  as the backend, which has a total bandwidth of 1 GHz and 16384 channels, providing a velocity 
resolution of $\sim$0.2 km s$^{-1}$ (see Table \ref{table_lines_teles}). Typical system temperatures are 200 K
and 120 K for the upper and lower sidebands respectively. The half-power beam width and the main beam efficiency 
($\eta$) are about 56\arcsec~(0.08 pc) and 0.5\footnote{\url{http://english.dlh.pmo.cas.cn/fs}},
respectively. 

\subsection{NRO 45 m}
We applied  OTF mapping observations of the  HCO$^+$ $J$ = 1--0, SO $J_N$ = 2$_2$--1$_1$, 
HC$_3$N $J$ = 11--10, $J$ = 10--9 and C$^{18}$O $J$ = 1--0 transitions towards  L1251-A  (Fig. \ref{herscmap}) 
using the NRO\footnote{The Nobeyama Radio Observatory is a branch of the National Astronomical Observatory of Japan, 
National Institutes of Natural Sciences. } 45 m telescope during 2019 May 8th to 14th. The size of the mapped region is 
25\arcmin$\times$10\arcmin. A four-beam receiver  \citep[FOREST;][]{2016SPIE.9914E..1ZM} 
was adopted as frontend and  SAM\,45 \citep{2012PASJ...64...29K} was adopted 
as backend. HCO$^+$ $J$ = 1--0, SO $J_N$ = 2$_2$--1$_1$, HC$_3$N $J$ = 11--10 and $J$ = 10--9 were observed 
simultaneously. The half-power beam width (HPBW) and beam efficiency of FOREST at 86 GHz were about 
19\arcsec~(0.028 pc) and 0.5, respectively. Spectra obtained with orthogonal linear polarizations were averaged to improve 
the signal-to-noise ratio (SNR). The system temperature varied from 150 K to 300 K depending on the zenith 
angles and weather conditions during the observations. The data were baseline extracted and gridded using the 
package NOSTAR\footnote{\url{https://www.nro.nao.ac.jp/~nro45mrt/html/obs/otf/exporte.html}} 
provided by the NRO. The spectra were smoothed to a velocity   resolution of $\sim$0.2 km s$^{-1}$ and gridded in 
10\arcsec~intervals. The resulting  noise levels in the main beam brightness temperature 
($T_{\rm mb}$) scale varied from  0.05 to 0.17 K for different lines, depending on  integration 
times and system temperatures.

\subsection{JCMT HARP}
We mapped L1251-A (Fig. \ref{herscmap}) employing the JCMT OTF mode in CO $J$ = 3--2, under the project 
ID M19BP057\footnote{\url{http://www.cadc-ccda.hia-iha.nrc-cnrc.gc.ca/en/search/?Observation.proposal.id=M19BP057&Observation.collection=JCMT}}. 
The HARP receiver was adopted. The observations were performed in November 14, 2019, with an on-source time 
of $\sim$30 minutes. The size of the mapped region is 20\arcmin$\times$10\arcmin. The half power beam width 
is approximately 14\arcsec. We smoothed and gridded the data cube with a pixel size 
15\arcsec$\times$15\arcsec~and a channel spacing of 0.2 km s$^{-1}$, to improve the signal-to-noise ratio.
The 1 sigma rms noise level is 0.5 K.

\section{Our targeted sources in the L1251 region} \label{sec_L1251}
The dark cloud L1251 \citep{1962ApJS....7....1L}   is located in the molecular ring 
of the Cepheus Flare and was found to be a site of recent low mass star formation 
\citep{1989ApJ...343..773S,1994ApJ...435..279S,1995PASP..107...49R,2015ApJS..218....5K}.

The distance is adopted as $300\pm 30$ pc derived from the 3-dimensional (3D) dust map based on Gaia 
\citep{2019ApJ...887...93G}. It is consistent with the values of 300$\pm$50 pc adopted by 
\citet{1993A&A...272..235K} and 330$\pm$30 pc  adopted by \citet{2004A&A...425..133B} and 
\citet{2019MNRAS.488..495W}. Based on CO observations, the L1251 cloud was divided into five substructures 
by \citet{1994ApJ...435..279S}, and these substructures were designated by letters from A to E (see the
top panel of Fig. \ref{herscmap}). The L1251 cloud is elongated in east-west direction, consisting of a 
dense ``head'' in the east and a diffuse ``tail'' in the west \citep{1989ApJ...343..773S,1994ApJ...435..279S}.

Four low-luminosity young stellar objects \citep{2008ApJS..179..249D} are located in the tail L1251-A, named 
as IRS1--4 from west to east. These were identified by the $c2d$ team based on the colors in Spitzer bands 
\citep{2008ApJS..179..249D,2009ApJS..181..321E,2009ApJ...692..973E}. The two sources, denoted as L1251-1 and 
L1251-2 in this work, contain IRS1 and IRS4, respectively (see Fig. \ref{herscmap}). The SCCC source denoted 
as L1251A (referred as L1251-IRS3 in this work) in \citet{2019MNRAS.488..495W} contains IRS3. 
A well-collimated jet in north-south direction 
associated with IRS3 (see also Sect. \ref{sec_jcmtoutflow}) can be explained with a small precession angle 
and a long stellar pulsating period \citep{2010ApJ...709L..74L}. L1251-3 is located in the head of L1251 (Fig. 
\ref{herscmap}), and it is also associated with two young stellar objects (YSOs) identified by Spitzer data.
Our targets L1251-1, L1251-2, and L1251-3 are all candidates of very low luminosity objects (VeLLOs) with 
$L_{int}<0.1\ L_\sun$, but may not be deeply embedded sources \citep{2008ApJS..179..249D}.

\subsection{YSO classfication and their continuum counterparts}
Continuum fluxes in Spitzer bands (3.6 $\mu$m, 4.5 $\mu$m, 5.8 $\mu$m, 8.0 $\mu$m, 24 $\mu$m and 70  $\mu$m), 
as well as other near infrared to milimetre continuum data with wavelengths ranging from 1.25 $\mu$m to 850 
$\mu$m were taken from the literature, and are listed in Table \ref{tab:4stars}.

A near-infrared 2MASS point source counterpart is found near IRS2 within 3\arcsec~(900 AU). All four YSOs associated with 
L1251-A have counterparts within 5\arcsec~(1500 AU) in Herschel's PACS Point Source Catalogue. The 70 $\mu$m 
fluxes of the four YSOs obtained by Herschel are about 1.2 times of the values obtained by Spitzer.  This 
difference may be introduced by different source extracting algorithms. IRS-3 and IRS-4 have been 
mapped by \citet{2007AJ....133.1560W} at 350 $\mu$m using SHARC-II mounted on the Caltech Submillimeter 
Observatory (CSO). Three SCUBA-2 cores identified by the JCMT Gould Belt Survey \citep{2017MNRAS.464.4255P}
(with core number 91, 82 and 57) are centered close  to ($<20\arcsec$) IRS2, IRS3 and IRS4, respectively.
These angular separations are comparable with the full width at half maximum (FWHM) source sizes  
($\sim$30\arcsec) and SCUBA beam sizes  ($\sim$10\arcsec~at 450 $\mu$m and $\sim$15\arcsec~at  850 $\mu$m). 
The SCUBA-2 core 95 is separated from L1251-3 by $\sim$ 1\arcmin. The SCUBA-2 core 71  
\citep{2017MNRAS.464.4255P} locates between IRS1 and IRS2, with IRS1 in its southwestern margin and IRS2 in 
the eastern margin. All four YSOs in L1251-A except IRS1 have  continuum detection at 1.2 mm  
\citep{2008A&A...487..993K} by the Max-Planck Millimetre Bolometer (MAMBO)  at the IRAM 30 m 
telescope.

These YSOs are classified based on color criteria and  spectral energy distribution (SED) fittings 
(Appendix \ref{sec_YSO}). IRS3 and IRS4 are classified as Class 0/I and Class II respectively. IRS1 and IRS2 
are classified as Class Flat. L1251-3 contains a Class II young stellar object.

\subsection{Large scale structures} \label{sec_LSS}
The map of the L1251 region from Herschel Space Telescope data consists of
hierarchical filaments and filamentary sub-structures (see Fig. \ref{herscmap}). Not only the tail of L1251, but also the ``head'' of 
L1251 can be visually divided into several filamentary sub-structures. To more accurately unveil the 
filamentary structures in this region, we obtained a surface density as well as a temperature map of the dust in 
the L1251 region from the Herschel images. The fitting procedures are described in Appendix \ref{pixbypix_sed} 
(see Fig. \ref{herscmap}).

\subsubsection{Extracting filaments} \label{sec:fil_ext}
The filamentary structure is extracted from the surface density map  using 
FilFinder\footnote{\url{https://github.com/e-koch/FilFinder}} (a python package developed by  
\citet{2015MNRAS.452.3435K}) to identify and analyse filaments. The skeleton of the extracted filamentary 
structures is shown in the lower panel of Fig. \ref{herscmap}. Six filamentary sub-structures, named as F1 
to F6 were identified in L1251 and L1247. L1251-A is associated with F1. The U-shaped ``head'' 
\citep{1994ApJ...435..279S} is made up by three distinguished sub-filaments, F2, F3 and F4. The southwestern 
ends of F2 and F3, together with the eastern end of F1, converge to a hub region that we 
called H (see the lower panel of Fig. \ref{herscmap}). The hub region is much less dense than the main parts 
of the three sub-filaments. 

The multiforked fibrous sub-structure F6 at the western side of L1251-A (see Fig. \ref{herscmap}), 
does not belong to L1251 but is a part of L1247. However, a slender and curved sub-filament F5 
with very low surface density (about one tenth of the values for the other sub-filaments) 
is found connecting F1 and F6. L1251-A (F1) is regarded as the western tail of the L1251 region by 
\citet{1994ApJ...435..279S} based on CO images. The F1-6 make up a dragon-like skeleton, in which F1 is the
spine instead of the tail of the whole structure consisting of  L1251 and L1247.

\subsubsection{Substructures in L1251-A} \label{sec_substructures}
In the L1251-A region, there is a side branch with a shape analogous to a letter ``S'' rotated counterclockwise 
by almost 90 degrees (Figs. \ref{herscmap} and \ref{fig:dust_hcopclumps}). We named this side branch 
intertwining with the main branch as F1-S. The segment of the main branch tangled by F1-S is denoted as F1-M, 
which contains the 4 YSOs in L1251-A. F1-M and F1-S make up a lying down ``8'' with two cavities. The eastern 
parts of F1-M and F1-S surrounding the eastern cavity (C1) are denoted as F1-ME and F1-SE, while the western 
parts surrounding the western cavity (C2) are denoted as F1-MW/F1-SW (See Fig. \ref{herscmap} and Fig.
\ref{fig:dust_hcopclumps} for the details).

\section{Parameters of the radio K-band Lines} \label{sec:singleobs}
The spectra  obtained in K-band using the Effelsberg 100 m telescope include HC$_3$N $J$ = 2--1, 
HC$_5$N $J$ = 9--8, 8--7 and 7--6, C$_3$S $J$ = 4--3, NH$_3$ (1,1), NH$_3$ (2,2), C$_4$H $N$ = 2--1 
$J$ = 5/2--3/2 $F$ = 3--2, c-C$_2$H$_2$   $1_{1,0}$--$1_{0,1}$ and c-C$_2$H$_2$   $2_{2,0}$--$2_{1,1}$. 
These spectra are shown in Fig. \ref{effel_example_spectra}. 

The lines observed in K-band are  fitted with single Gaussian profiles, except for NH$_3$ (1,1), which has been fitted
with the hyperfine structures (HFS; see Sect. \ref{sec4_1}). We also tried to apply HFS fitting to HC$_3$N $J$ = 2--1, 
but excitation temperatures are not available because HC$_3$N $J$ = 2--1 is optically thin in most sources
(Sect. \ref{sec4_1}; see also \citet{2013A&A...553A..58L}). The hyperfine structure of HC$_3$N $J$ = 2--1, 
except for the blended $F$ = 3--2 and  $F$ = 2--1 features, is only marginally detected in L1251-2, and not detected 
in L1251-1 and L1251-3. The peak optical depth  of the unsplit HC$_3$N $J$ = 2--1 (assuming the central 
frequencies of all hyperfine structures are aligned) of L1251-2 is $0.20\pm 0.15$, given by the standard CLASS 
HFS\footnote{\url{https://www.iram.fr/IRAMFR/GILDAS/}} procedure. We conclude that the HC$_3$N $J$ = 2--1 lines 
are optically thin for the observed sources. The fitted parameters are listed in Table \ref{tab:Kband_fit}.

The column densities of H$_2$ ($N^{dust}$(H$_2$)) and dust  temperatures ($T_{\rm dust}$) of L1251-1, L1251-2 and 
L1251-3 are extracted from the surface density and dust temperature maps (see Sect. \ref{sec_LSS}), and are smoothed to 
the 40\arcsec~beam size of the Effelsberg telescope. The derived $N^{dust}$(H$_2$) and  $T_{\rm dust}$ are listed in 
Table \ref{tab:derivedpar1}.

\subsection{NH$_3$ and linear CCMs} \label{sec4_1}
Optical depths ($\tau$), excitation temperatures ($T_{\rm ex}$) of NH$_3$ (1,1),  rotational temperatures 
($T_{\rm rot}$) and column densities of NH$_3$, as well as the volume density of H$_2$ derived from NH$_3$ 
emissions ($n^{NH_3}$(H$_2$)) have been calculated (see Appendix \ref{sec_nnh3}). The fitted line parameters 
of NH$_3$  are listed in Table \ref{tab:NH3_pars}. The derived parameters for NH$_3$, $T_{\rm ex}$(NH$_3$), 
$T_{\rm rot}$(NH$_3$) and $n^{NH_3}$(H$_2$)  are listed  in Table \ref{tab:derivedpar1}.

All the species HC$_3$N, HC$_5$N, C$_3$S and C$_4$H are linear carbon-chain molecules. To calculate the column 
densities, excitation temperatures should be known. Unfortunately, the three rotational lines of HC$_5$N can 
not be used to derive the excitation temperatures $T_{\rm ex}$, since they all have upper level energies lower 
than 6 K. The HC$_5$N $J$ = 9--8 line has been adopted to calculate the column density of HC$_5$N, since it
has the highest SNR. The excitation temperatures are assumed  to be identical with the rotational temperatures 
of NH$_3$, which are close to the dust temperatures, with differences smaller than 1 K. The column densities 
of these carbon-chain species are calculated under the assumption of local thermal equilibrium (LTE), using Eq. 
\ref{eq:col_general} in Appendix \ref{sec_nnh3}, with the line strength $S$ and partition function $Q_{\rm rot}$ quoted from 
Splatalogue\footnote{\url{http://www.cv.nrao.edu/php/splat/}}. If the excitation temperature is changed by 2 K, 
the calculated column density will change by less than 20 percent. 
The dipole moment of C4H is adopted as 0.9 D. If the value of 2.10 D is adopted \citep{2020ApJ...890...39O} 
the evalued column densities will be a factor of 6 smaller than the presented ones. 

The column densities and abundances of NH$_3$ and CCMs are listed in Table \ref{tab:derivedpar}.

\subsection{c-C$_3$H$_2$} \label{sec_c3h2}
Cyclopropenylidene (c-C$_3$H$_2$), the first detected interstellar organic ring molecule, is a typical 
constituent of the dense interstellar medium \citep{1985ApJ...299L..63T,1985ApJ...298L..61M,1987ApJ...314..716V}.
c-C$_3$H$_2$ has $C_{2v}$ symmetry. Two H atoms are coupled to generate the $ortho$ 
(nuclear spin = 1) and $para$ (nuclear spin = 0) species of c-C$_3$H$_2$ with spin statistical weights of 3 and 
1, respectively \citep{2006A&A...449..631P}. Because of the Pauli exclusion principle, $para$ c-C$_3$H$_2$ is 
characterized by even $K_a+K_c$, and $ortho$  c-C$_3$H$_2$ by odd $K_a+K_c$ values. The abundance ratios between 
the $ortho$ c-C$_3$H$_2$  (o-C$_3$H$_2$) and $para$ c-C$_3$H$_2$ (p-C$_3$H$_2$) in TMC-1 are all larger than one, 
and can reach 3 for  evolved cores. The $ortho$-to-$para$ ratio (o/p ratio) of c-C$_3$H$_2$ is 3.1$\pm 0.4$ 
in L1257 \citep{2001PASJ...53..251T,2015ApJ...807...66Y}. \citet{2006A&A...449..631P}  modeled the o/p ratios 
of c-C$_3$H$_2$, and found that the o/p ratios of c-C$_3$H$_2$ can reach 3 if exchange processes involving 
H$^+$ and protonating ions HX$^+$  were considered, even if H$_2$ (the precursor of c-C$_3$H$_2$)
is overwhelmingly $para$.

Fig. \ref{c3h2_diag} in Appendix \ref{c3h2_model} 
shows the energy diagram of c-C$_3$H$_2$. The states 2$_{2,0}$ and 2$_{1,1}$ of $para$ c-C$_3$H$_2$ 
have similar upper-level energies ($\sim 9$ K). The transition between 2$_{1,1}$ and 1$_{1,1}$ is 
forbidden\footnote{\url{https://home.strw.leidenuniv.nl/~moldata/}}, and c-C$_3$H$_2$ in state 2$_{1,1}$ can only 
transit to $2_{0,2}$, with a spontaneous emission coefficient 2.679$\times 10^{-6}$ s$^{-1}$.
However, the spontaneous emission coefficient between 2$_{2,0}$ and $1_{1,1}$ is large (5.354$\times 10^{-5}$ s$^{-1}$).
Thus the state 2$_{2,0}$ is more difficult to populate than $2_{1,1}$, and the population ratio between 2$_{2,0}$ 
and $2_{1,1}$ is small with excitation temperatures lower than the temperature of the background radiation.
c-C$_3$H$_2$ in state 2$_{1,1}$ will be pumped to $2_{2,0}$ through absorption of background photons (e.g. 
the cosmic microwave background), and further transit to 1$_{1,1}$. This is the reason why c-C$_3$H$_2$ 2$_{2,0}$--2$_{1,1}$ 
shows absorption features in all of our detected sources. With a typical H$_2$ volume density  $n=10^5$ cm$^{-3}$,
kinetic temperature $T_{\rm kin}=10$ K and c-C$_3$H$_2$ column density $N=10^{13}$ cm$^{-2}$, the excitation 
temperatures of c-C$_3$H$_2$ 2$_{2,0}$--2$_{1,1}$  and 1$_{1,0}$--1$_{0,1}$ from RADEX are 1.2 K and 5.4 K, respectively. 

We assume a constant o/p ratio of 3 here to fit the c-C$_3$H$_2$ 1$_{1,0}$--1$_{0,1}$ (ortho type) emission 
and the c-C$_3$H$_2$ 2$_{2,0}$--2$_{1,1}$ (para type) absorption lines, following the fitting method described in 
Appendix \ref{c3h2_model}. The $\mathcal{R}$ values (the ratio between the intensity of $para$ c-C$_3$H$_2$ 
2$_{2,0}$--2$_{1,1}$ and that of $ortho$ c-C$_3$H$_2$ 1$_{1,0}$--1$_{0,1}$) for L1251-1, L1251-2 and L1251-3 are
$-0.38$, $-0.22$ and $-0.40$, respectively. The volume densities ($n^{c-C_3H_2}$(H$_2$)) of L1251-1 and L1251-3 
can not be well constrained through Eq. \ref{eq_R_rough}  
in Appendix \ref{c3h2_model}  and the values are adopted as  10$^4$ cm$^{-3}$, the value
below which the $\mathcal{R}$ value will be little affected by $n$.
The volume density of L1251-2 is estimated as $3\times 10^5$ cm$^{-3}$ from $\mathcal{R}$  
using Eq. \ref{eq_R_rough}.
However, we still allow volume density to vary and fit the column density of c-C$_3$H$_2$ and volume density simultaneously. 
The kinetic temperatures are adopted as $T_{\rm rot}$(NH$_3$) for 
the fittings (Appendix \ref{c3h2_model}). The column densities of c-C$_3$H$_2$  for L1251-1, L1251-2 and L1251-3 given 
by the fittings are 6.7$\times$10$^{13}$, 1.7$\times$10$^{13}$ and 4.8$\times$10$^{13}$ cm$^{-2}$, respectively.   
The fitted $n^{c-C_3H_2}$(H$_2$) is 2.2$\times$10$^5$ cm$^{-3}$ for L1251-2. It should be noted that the emission model 
described in Appendix \ref{c3h2_model} assumes that the emissions are from a uniform medium. The column densities of 
c-C$_3$H$_2$ for L1251-1 and L1251-3 will be over-estimated because of the under estimations of the volume densities. 
The c-C$_3$H$_2$ 1$_{1,0}$--1$_{0,1}$ emission may mainly originate from the inner denser regions, 
while  the absorption features of c-C$_3$H$_2$ 2$_{2,0}$--2$_{1,1}$ may originate from the outer  diffuse 
regions, where these molecules may also be quite abundant \citep{1989AJ.....97.1403M} but can not be effectively populated.

\subsection{Signposts of SCCC}
L1251-1 and L1251-3 both have integrated intensities of C$_3$S $J$ = 
4--3 larger than those of HC$_5$N $J$ = 9--8. In contrast,  L1251-2 
shows much weaker emission of C$_3$S $J$ = 4--3 compared with HC$_5$N $J$ = 9--8. Since 
C$_3$S and HC$_5$N are both linear molecules with similar electric dipole moments (3.6 and 4.33 respectively), 
their abundance ratio can be approximately represented by the integrated intensity ratio between their lines at 
similar frequency. This criterion can  help us quickly find candidate sources with N(C$_3$S)$>$N(HC$_5$N).
The abundance ratios between C$_3$S and HC$_5$N are 2.1$\pm$0.4 and 1.2$\pm$0.2 for L1251-1 and L1251-3, but 0.4$\pm$0.1 
for L1251-2. Such high abundances of C$_3$S in L1251-1 and L1251-3 are quite rare in cold clouds and 
have not been measured in star-forming regions before (see Sect. \ref{abnornalc3s}). 

\begin{figure*}
\centering
\includegraphics[width=0.4\linewidth]{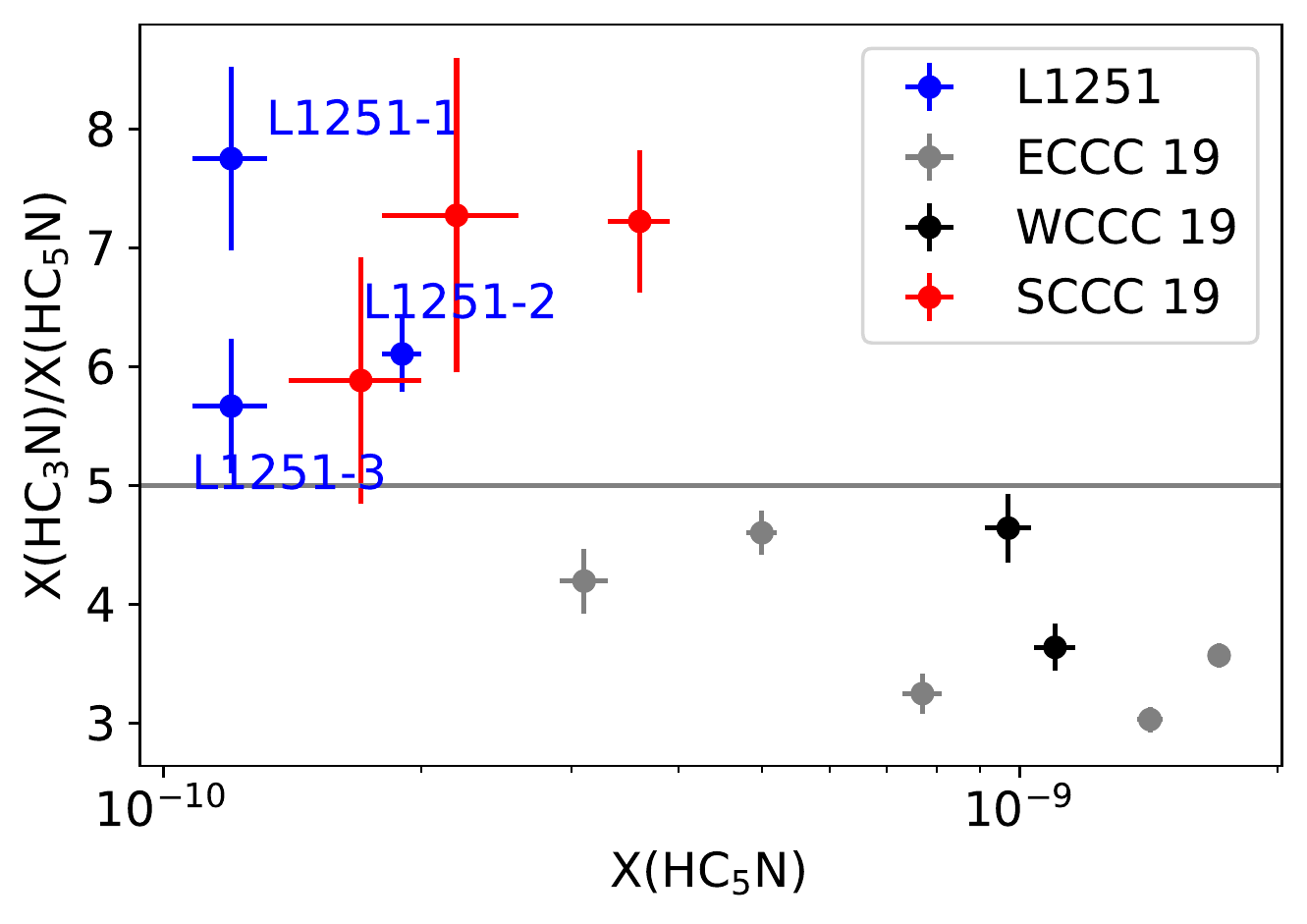}
\caption{The ratio between the column densities of HC$_3$N and HC$_5$N.
The three blue points represent L1251-1, L1251-2 and L1251-3.
Red and gray points show the data for SCCC sources and early carbon-chain cores (ECCC) from 
\citet{2019MNRAS.488..495W}, detected in Ku band. 
Black points represent WCCC sources Lupus I-1 (\citep[IRAS 15398-3359]{2009ApJ...697..769S,2019MNRAS.488..495W})  
and L1489 EMC \citep{2019A&A...627A.162W}.
\label{HC3N_HC5N_fig}}
\end{figure*}

For L1251-2 and L1251-3, the column density ratios between HC$_3$N and HC$_5$N ($N$(HC$_3$N)/$N$(HC$_5$N))  are
about 4, similar to the value of starless cores and common outflow sources 
\citep{1998A&A...329.1156T,2012MNRAS.419..238B,2016ApJ...817..147T,2018ApJ...863...88L}. L1251-1 has a larger 
$N$(HC$_3$N)/$N$(HC$_5$N)  ratio  of 8$\pm$1 (see Fig. \ref{HC3N_HC5N_fig}), which is closer to the values of 
other SCCC sources \citep{2019MNRAS.488..495W}.

Abundant C$_4$H, especially derived from its high excitation lines with $\Delta E_u>20$ K, is the key to study 
the heating of the ambient medium by YSOs and to investigate the warm carbon chain  chemistry , as suggested
by \citet{2008ApJ...672..371S,2009ApJ...697..769S}. From our observations, L1251-1 and L1251-3 have larger abundance 
ratios between C$_4$H and HC$_3$N (9.0 and 8.6) than that of L1251-2 (4.5). The sources  
detected in L1251 have similar column densities of C$_4$H compared to the WCCC source L1527 
(8.8$\pm$0.5 $\times$ 10$^{13}$ cm$^{-3}$). 
The upper level energy of the detected  C$_4$H line is low (see Table \ref{table_lines_teles}), and a contribution 
of the emission from an extended component can not be entirely ruled out.  Nevertheless, the WCCC theory is  
a possible explanation for the large hydrocarbon abundances, as suggested by  \citet{2011ApJ...730L..18C} based on 
their observations near L1251-IRS3. However,  their measured position is one arcmin farther away from the stellar source than the 
one observed by \citet{2019MNRAS.488..495W}, which shows that the precursors of the shocked region in L1251-IRS3  may be WCCC regions.
The abnormally high abundances of C$_3$S we detected imply an influence of shocks in the parent gas component 
with high abundances of hydrocarbons.

For L1251-1 (IRS-1), the line width of C$_3$S $J$=4--3 (2.5 km s$^{-1}$) is obviously higher than the values 
($<$1.4 km s$^{-1}$) of  transitions of other species except C$_4$H (see Table \ref{tab:NH3_pars}), even  considering the 
low velocity resolution ($\sim$0.5 km s$^{-1}$) of the observed spectra. This large line width is introduced by the red 
velocity component in the C$_3$S spectrum (see Fig. \ref{effel_example_spectra}), which we suggest as an evidence of 
outflow activity driven by IRS-1 (see also Sect. \ref{sec_jcmtoutflow}). The relatively large line width of C$_4$H 
(1.8 km s$^{-1}$) may also be affected by star formation activities.

Overall, L1251-1  characterized by $N$(C$_3$S)$>$$N$(HC$_5$N) is  a clear SCCC source
that were looking for.
L1251-3 has $N$(C$_3$S) slightly larger than $N$(HC$_5$N) but lacks of 
mapping observagtions and is identifed as a candidate SCCC source. 

\section{Mapping analysis towards L1251-A}\label{sec:mapping}
\subsection{Outflows traced by CO $J$ = 3--2 lines}\label{sec_jcmtoutflow}
Features of outflows are pronounced in the L1251-A region traced by CO $J$ = 3--2 as shown in Fig. \ref{jcmt_co}.
\subsubsection{Outflows from IRS3}
A bipolar CO $J$ = 3--2 jet  originating from IRS3 extending in north-south direction  has been detected.
The blue part of the jet consists of two bullets. The bullet close to IRS3  corresponds to the slender
blue finger in the position-velocity (P-V) map (near the zero offset of the lower right panel of Fig. 
\ref{jcmt_co}), and the other one located more than 2\arcmin~(0.17 pc) away from IRS3 corresponds to a 
blueward convex bulge (see the lower right panel of Fig. \ref{jcmt_co}). The jet traced by CO $J$ = 3--2 is 
similar to that previously revealed by a CO $J$ = 2--1 image \citep{2010ApJ...709L..74L}. A jet with a small 
precession angle and long pulsating period was introduced by \citet{2010ApJ...709L..74L} to explain their 
observations.

Among the four YSOs in L1251-A, only IRS3 is accompanied by characteristics of strong bipolar jets in Spitzer maps. 
From the CO $J$ = 3--2 observation,   the red and blue lobes are spatially 
separated by 6\arcmin~with a velocity separation of about 8 km s$^{-1}$. The derived dynamical timescale  is about
$6\times 10^4$ yr, which is consistent with the value $5\times 10^4$ yr 
given by \citet{2010ApJ...709L..74L}. This outflow activity was thought to be the driving source of SCCC around IRS3 
\citep{2019MNRAS.488..495W}. The high abundance of C$_3$S relative to HC$_5$N in L1251-1 (IRS1), may originate 
from outflows driven by IRS1. The outflow activity has extinguished now,  since in our model 
\citep{2019MNRAS.488..495W},  the enhancement of C$_3$S will not lead to X[C$_3$S]$>$X[HC$_5$N] till 
10$^5$ yr after the start of the enrichment of the gas phase  by sulfur 
from shocks induced by outflows.
The SCCC sources found in \citet{2019MNRAS.488..495W} all have outflows with timescales smaller than 10$^5$ yr, 
and are thus in an earlier stage of SCCC with $X$[C$_3$S]$>$$X$[HC$_7$N] but $X$[C$_3$S]$<$$X$[HC$_5$N].

\subsubsection{Outflows from IRS4}
\begin{figure*}
\centering
\includegraphics[width=0.7\linewidth]{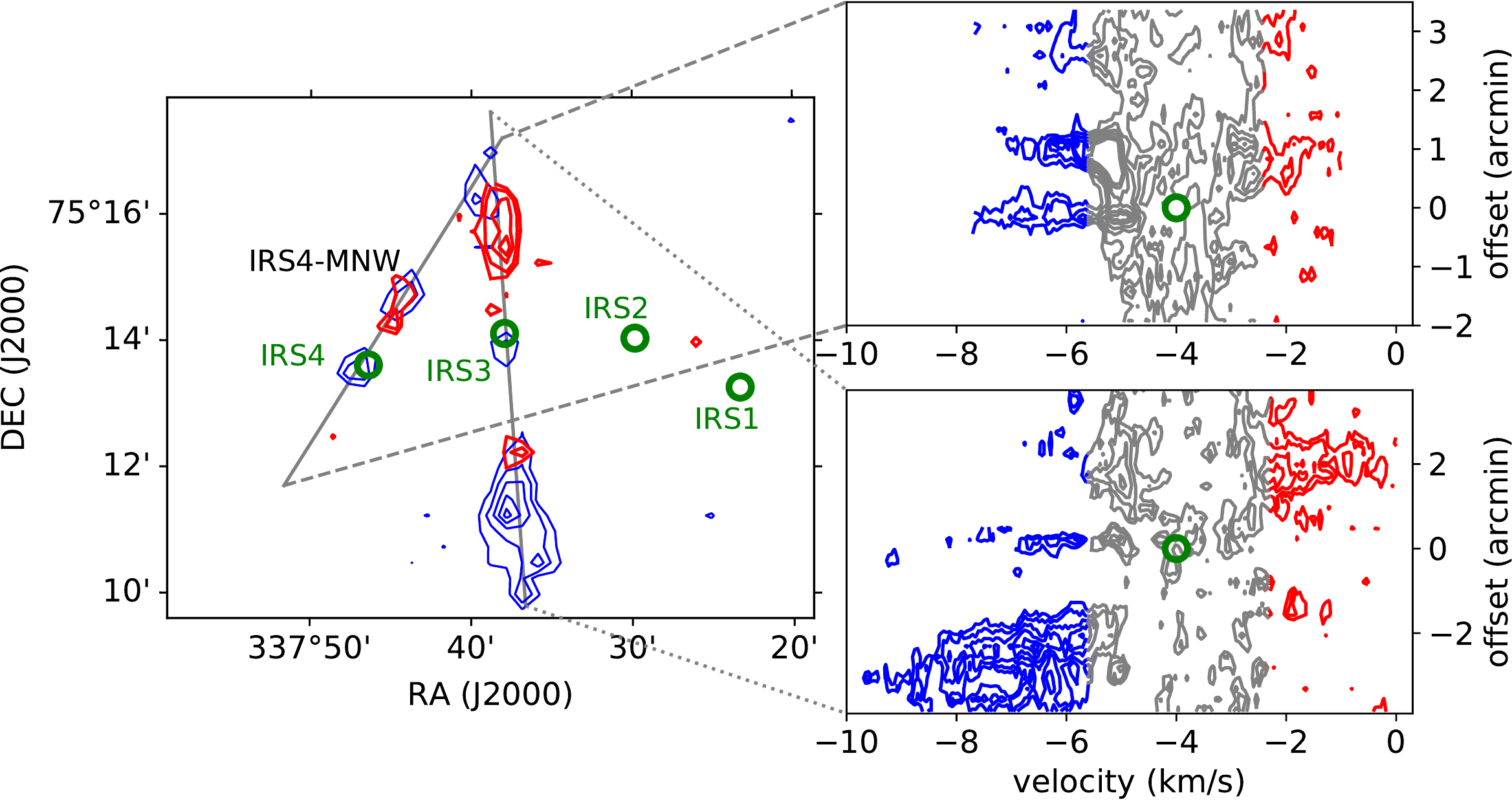}
\caption{Left: Red and blue contours represent the  integrated intensities of the red 
(from $-2.5$ km s$^{-1}$ to $-1$ km s$^{-1}$) and blue (from $-9$ km s$^{-1}$ to $-5.5$ km s$^{-1}$) wings
of the JCMT CO $J$ = 3--2 lines. Green circles mark
the infrared sources IRS1 to IRS4. 
Right: Position-velocity maps.
IRS4 and IRS3 have zero offsets in the upper and lower right panels, respectively. 
The contour levels are from 25 percent to 85 percent with an increment of 15 percent relative to the maximum intensities. 
\label{jcmt_co}}
\end{figure*}

A blue lobe in south-east direction is detected in CO $J$=3--2 around IRS4 (L1251-2; see Fig. \ref{jcmt_co}). 
It is more compact than that detected in
CO $J$ = 2--1 using the six-meter telescope of the Seoul Radio Astronomy Observatory (SRAO)  \citep{2019ApJS..240...18K}. 
However, a redward instead of the blueward lobe was also detected towards IRS4  
using the SRAO \citep{2010ApJ...709L..74L}. In our P-V map (upper right panel of Fig. \ref{jcmt_co}), 
a blue finger at an offset of +1\arcmin~is associated with IRS4. A similar but
red finger at the same offset indicates a small angle between
the direction of the jet and the line of sight. This may explain why the blue and red lobe are coincident. 
It should be noted that a small red component with a velocity range from -1 km s$^{-1}$ to 0 km s$^{-1}$ can 
be seen from the spectra of CO $J$ = 2--1 \citep{2019ApJS..240...18K}, and it may contribute in a non-ignorable 
way to the intensity of the red lobe integrated from -1 km s$^{-1}$ to 5 km s$^{-1}$ by \citet{2010ApJ...709L..74L}.

\begin{figure*}
\centering
\includegraphics[width=0.45\linewidth]{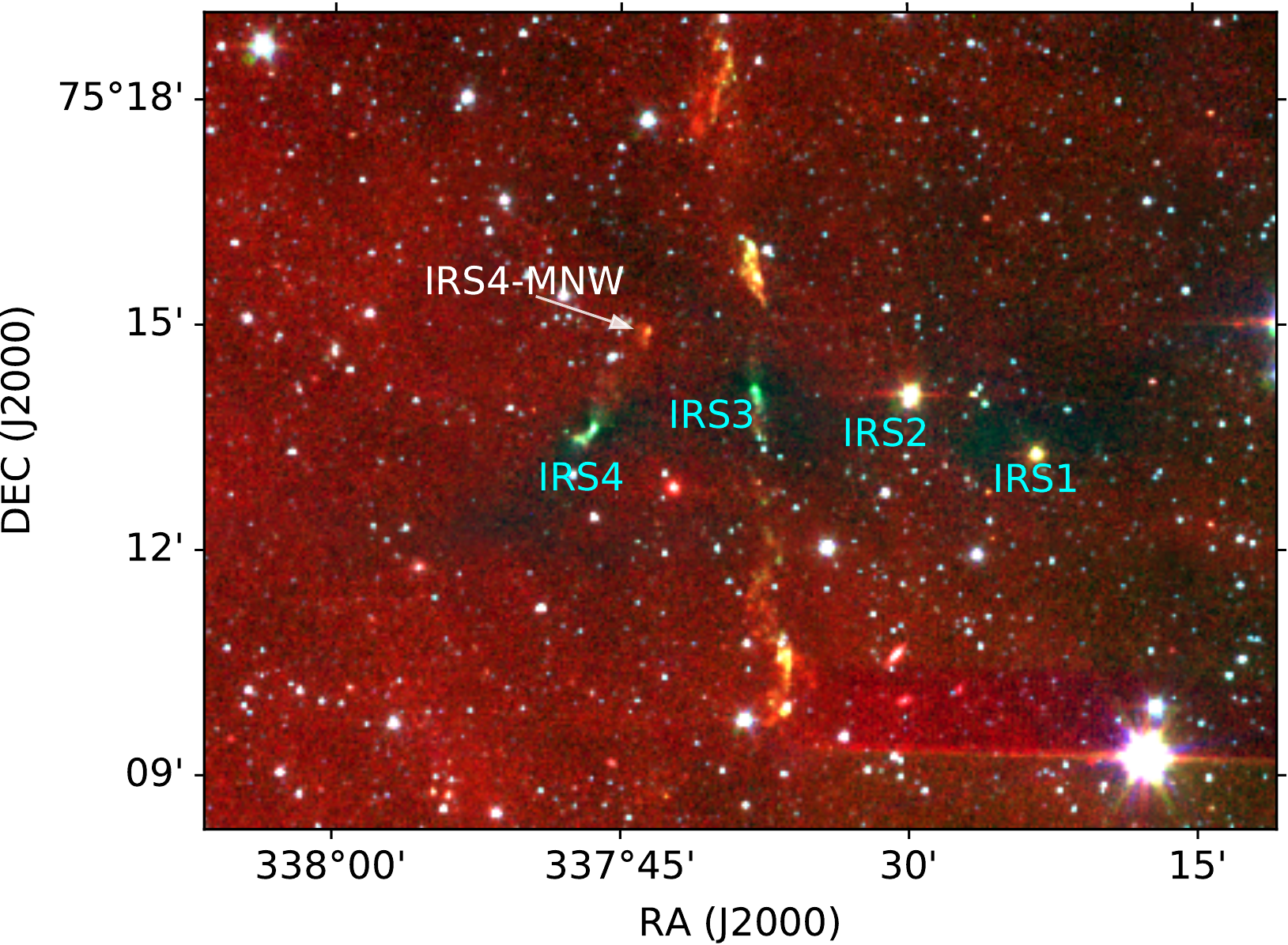}
\caption{Three-color image comprised of Spitzer IRAC 3.6 (blue), 4.5 (green), and 8.0 (red) $\mu$m images.\label{spitzer_map}}
\end{figure*}

The clump located in the northwest of IRS4, denoted as IRS4-MNW in the left panel of Fig. \ref{jcmt_co}, 
is accompanied by both blue and red lobes. These two lobes correspond to the blue and red fingers in 
the P-V map (upper right panel of Fig. \ref{jcmt_co}) with an offset of approximate 1\arcmin. The three-color image 
comprised of IRAC 3.6, 4.5, and 8.0 $\mu$m images is shown in Fig. \ref{spitzer_map}. IRS4-MNW is associated with 
a star-like object at the northwestern edge (22:30:54 +75:14:54), with an infrared spectral index $\alpha_{IR}$ 
(defined in Appendix \ref{sec_YSO}) of 1.24 \citep{2009ApJS..181..321E}.
This object was not classified as a  young stellar object because it has a weak Spitzer/MIPS 24 $\mu$m flux (0.212 mJy) 
and was not detected in the Spitzer/MIPS 70 $\mu$m band. Its color and morphology shown in Fig. \ref{spitzer_map} are 
similar to the shocked regions driven by IRS3. This object may represent shocked gas. 

\subsubsection{Hint for outflows from L1251-1 (IRS1)}
\begin{figure*}
\centering
\includegraphics[width=0.27\linewidth]{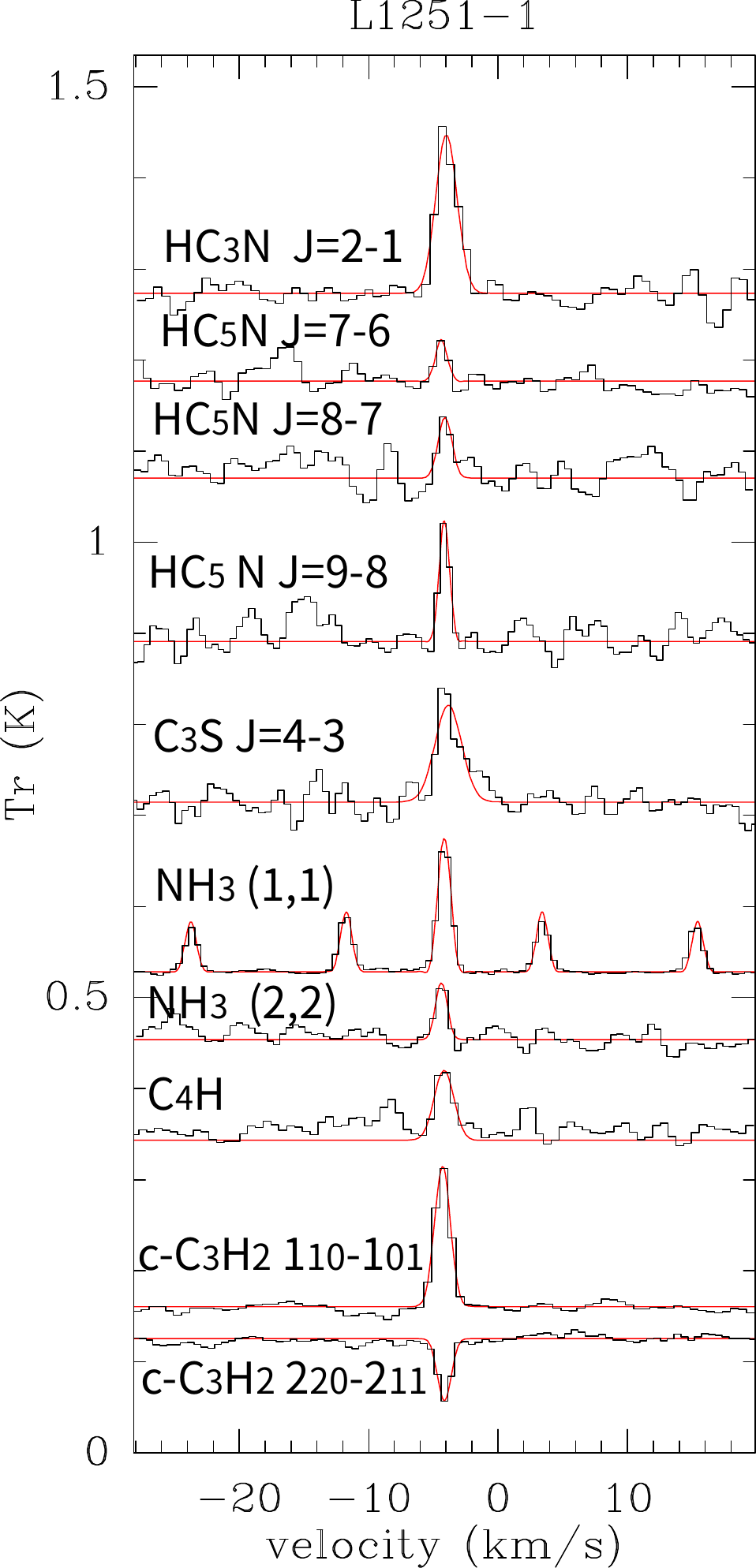}
\includegraphics[width=0.27\linewidth]{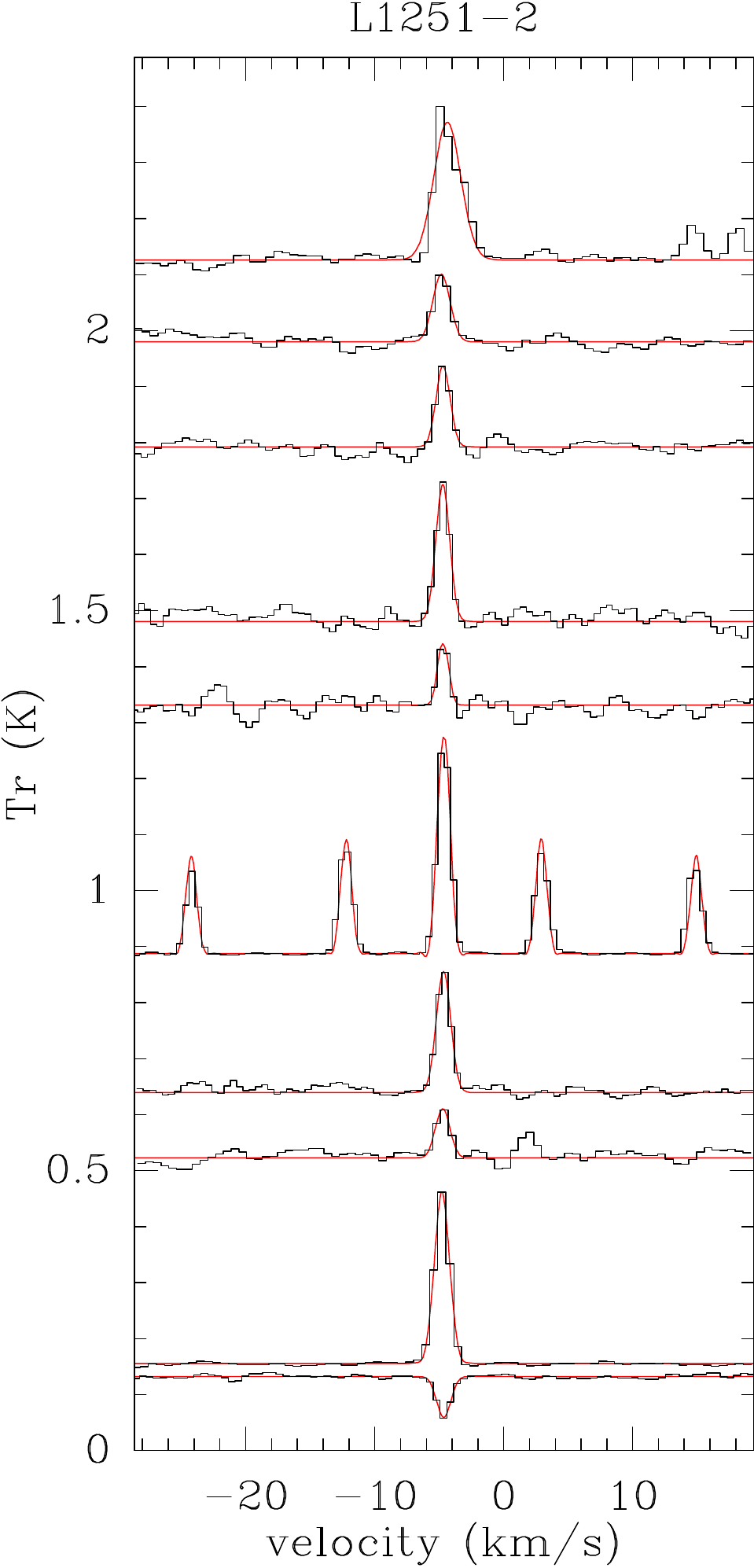}
\includegraphics[width=0.28\linewidth]{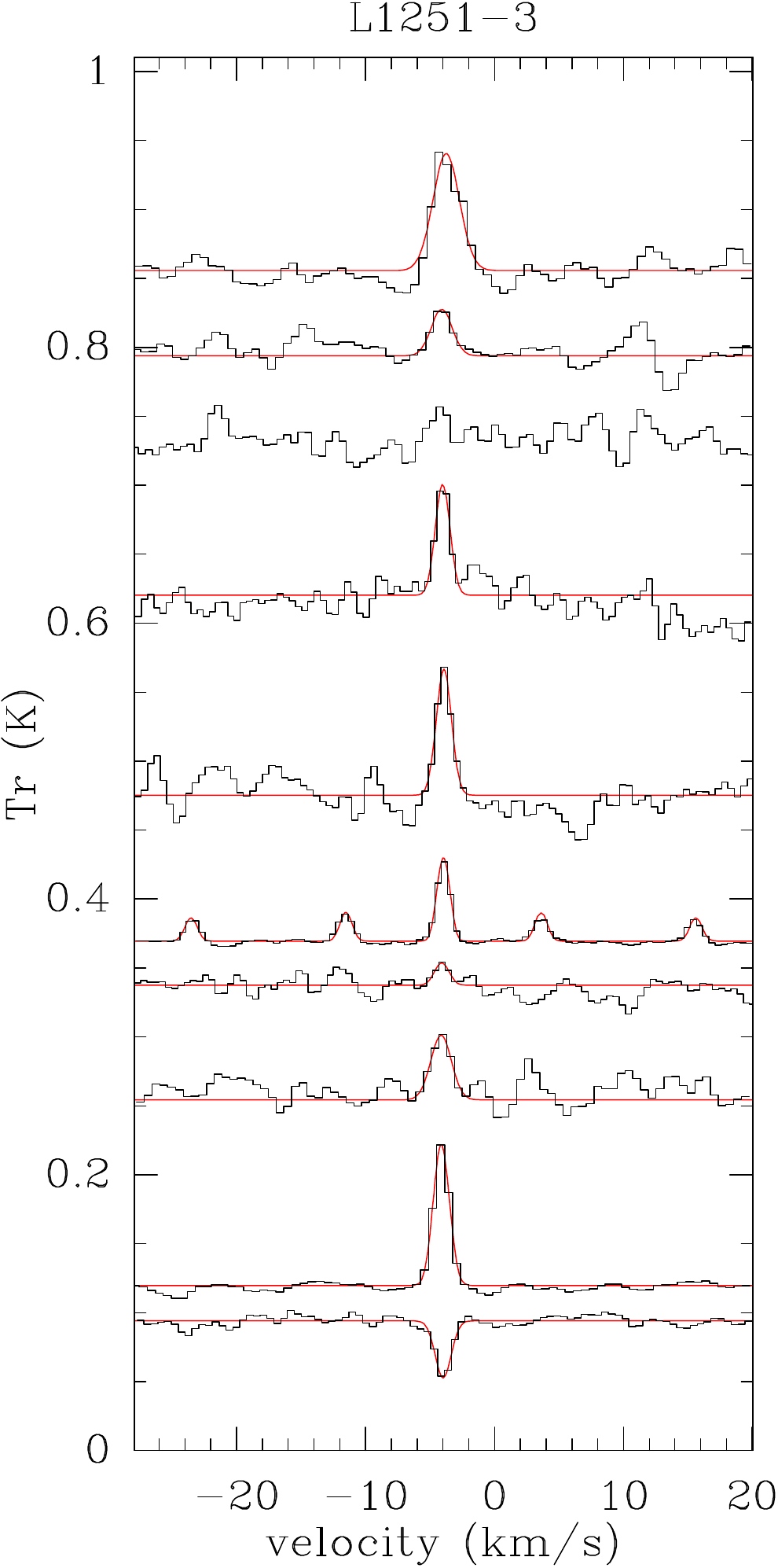}
\caption{The  K-band spectra observed with the Effelsberg 100m telescope.
For each panel, the spectra from top to bottom are HC$_3$N $J$ = 2--1, HC$_5$N 
$J$ = 7--6, 8--7 and 9--8, 
C$_3$S $J$ = 4--3, NH$_3$ (1,1), NH$_3$ (2,2),
C$_4$H $N$ = 2--1, $J$ = 5/2-3/2, $F$ = 3--2, 
c-C$_2$H$_2$   $1_{1,0}$--$1_{0,1}$ and c-C$_2$H$_2$   $2_{2,0}$--$2_{1,1}$. 
The spectra of NH$_3$ (1,1), NH$_3$ (2,2), c-C$_2$H$_2$   $1_{10}$-$1_{01}$ and c-C$_2$H$_2$   $2_{2,0}$--$2_{1,1}$
have been divided by  factors of 10, 2, 3 and 3 respectively. \label{effel_example_spectra} }
\end{figure*}

Blue and red wing emissions were detected towards IRS1
in $^{12}$CO, although at a much weaker level than that in IRS3 \citep{2012AJ....144..115S}. 
The signature of the red C$_3$S $J$ = 4--3 wing of 
L1251-1 shown in Fig. \ref{effel_example_spectra} is also consistent with the redder velocity of IRS1 compared 
to IRS2 and IRS4 shown in the P-V diagram (Fig. \ref{fig:pv}). No wing signature of C$_3$S $J$ = 4--3 is detected 
in L1251-2 and L1251-3. The red C$_3$S $J$ = 4--3 wing in IRS1 extends from $-3$ km s$^{-1}$ to 0 km s$^{-1}$ 
and  is consistent with the idea that C$_3$S in this region is shock related.

Bipolar jets should be common in this region, and there might also exist episodic jets originating from IRS1 and IRS2, 
but these putative jets, if they existed at all, are extinguished now.

\subsection{Maps of the PMO 13.7m}
The integrated intensity maps of CS $J$ = 2--1, C$_2$H $N$ = 2--1  $J$ = 3/2--1/2 $F$ = 2--1 and N$_2$H$^+$ $J$ = 1--0 
$F_1$ = 2--1 are shown in Fig. \ref{fig:pmo_c2hn2hp} with maximum  values  of 0.53 (0.03), 
0.29 (0.02) and 0.88 (0.04) K km s$^{-1}$ around IRS4, respectively. The emission regions of C$_2$H and  
N$_2$H$^+$ are more compact than those of CS, with emission centers shifted northwards. Their shapes are convex and 
well correlated with the main F1 branch, which runs across the four YSOs. It is consistent with the fact that the 
northern part of L1251-A is denser (see Fig. \ref{fig:dust_hcopclumps}) and may be more evolved, since N$_2$H$^+$ 
tends to be enhanced in evolved regions \citep{2002ApJ...572..238C,2019A&A...622A..32L}, while CS $J$ = 2--1 
is usually characterized by more extended emission, which is getting closer to the size of the CO emission
region \citep{1993Ap&SS.200..183W}. Besides,  the abundance of CS tends to be enhanced 
in turbulent regions, and CS emission is  consistent with the chemical effects expected in shocked regions 
\citep{1980ApJ...236..182H,1989ApJ...346..168Z, 1992ApJ...392..551S,2019ApJ...885...82L}. The depletion of the CS 
molecule \citep{2018MNRAS.478.5514V,2020ApJ...891..169K} in the dense region of F1-M will also make the CS emission 
look more extended.  

Assuming the emission lines are optically thin and the excitation temperatures are the 
same as the dust temperatures of the dense regions,  $\sim$10 K (see Fig. \ref{herscmap} and Table 
\ref{tab:derivedpar1}),
the conversion factors between the line intensities and column densities are 4.7$\times$10$^{12}$, 
5.1$\times$10$^{13}$ and 2.4$\times$10$^{12}$ cm$^{-2}$ per K km s$^{-1}$ for CS $J$ = 2--1, 
C$_2$H $J$ = 3/2--1/2 $F$ = 2--1 and  N$_2$H$^+$ $J$ = 1--0 $F_1$ = 2--1, respectively.
The ratios between the column densities of C$_2$H and  N$_2$H$^+$ 
(X[C$_2$H]/X[N$_2$H$^+$]) have a mean value of 8 with a standard deviation of 3. The abundance ratios of  
N[C$_2$H]/N[N$_2$H$^+$] in the L1251-A region are comparable with the values of typical dark clouds and 
star-forming regions, but tend to be lower than the typical value ($>$10) of Planck Galactic cold clumps (PGCCs)
\citep{2019A&A...622A..32L}. In starless cores the abundance ratio between C$_2$H and N$_2$H$^+$ will 
decrease as the cores evolve, but it is not well
constrained in star forming regions. However, the low values of 
N(C$_2$H)/N(N$_2$H$^+$) in L1251-A imply that this region is to overall not in chemically young phase.

\subsection{Maps of the NRO 45m} \label{sec_45m}
\subsubsection{SO and C$^{18}$O}
The two cavities (C1 and C2) enclosed by the two twisted sub-filaments can also be identified clearly in the 
emission map of SO $J_N$ = 2$_2$--1$_1$ (see Fig. \ref{map:int_SO}), observed using the NRO 45 m. The dust emission 
from  F1-ME is much stronger than that from F1-SE, while F1-SE shows stronger SO $J_N$ = 2$_2$--1$_1$ emission. 
The C$^{18}$O emission region shows a similar morphology to SO $J_N$ = 2$_2$--1$_1$, except that the 
differences of the emission intensities of  C$^{18}$O between F1-SE and F1-ME are more notable than in SO. There 
are no C$^{18}$O clumps around IRS3 and IRS4. F1-ME, with  $V_{\rm LSR}$ of about $-3.8$ km s$^{-1}$, is slightly redder than 
F1-SE, with $V_{\rm LSR}$ about $-4.2$ km s$^{-1}$. The emission of SO and C$^{18}$O is  enhanced in F1-MW 
around IRS2, located in the intersection region between F1-M and F1-S. This may be caused by the feedback of IRS1 and 
IRS2, or by the dynamical interaction between F1-M and F1-S.

\subsubsection{HCO$^+$}
Gaussian fittings were applied pixel-by-pixel  to the spectra of HCO$^+$ $J$ = 1--0, using the Python 
package PySpecKit\footnote{\url{https://pythonhosted.org/pyspeckit/}}. The resulting maps of integrated 
intensities ($\mathcal{F}$), central velocities ($V_{\rm LSR}$) and line widths ($\Delta V$) are shown
in Fig. \ref{fig:hcop_gpar}. The fitted parameters of a pixel are kept only if the pixel itself and at least 
four of its eight adjacent pixels have SNRs larger than 5. The emission of HCO$^+$ $J$ = 1--0 around F1-ME 
overwhelms those around F1-SE, in contrast to the case for SO $J_N$ = 2$_2$--1$_1$. The typical line width of 
HCO$^+$ $J$ = 1--0 is near 0.6 km s$^{-1}$, except for the regions around the four YSOs and the margins 
of the emission region where the line widths of  HCO$^+$ $J$ = 1--0 can be larger than 1 km s$^{-1}$.

We have used the FellWalker \citep{2015A&C....10...22B}  source extraction algorithm, implemented as part of the
Starlink\footnote{\url{http://starlink.eao.hawaii.edu/starlink/WelcomePage/}} suite, to identify compact clumps 
from the HCO$^+$ map,  following the source extraction processes described in \citet{2015MNRAS.453.4264M} and 
\citet{2017MNRAS.469.2163E}. Fourteen HCO$^+$ dense clumps were identified, and they are shown in 
Fig. \ref{fig:dust_hcopclumps}. The peak positions of HCO$^+$ and dust clumps are misaligned. HCO$^+$ tends to 
assemble around the junction points of filamentary branches, e.g. H3, H9 and H11-H14. In contrast, the dust 
emission is more uniformly distributed along the main branch (Sect. \ref{sec_substructures}) of the filamentary 
structures (see Fig. \ref{fig:dust_hcopclumps}).

\subsubsection{HC$_3$N}
\begin{figure*}
\centering
\includegraphics[width=0.6\linewidth]{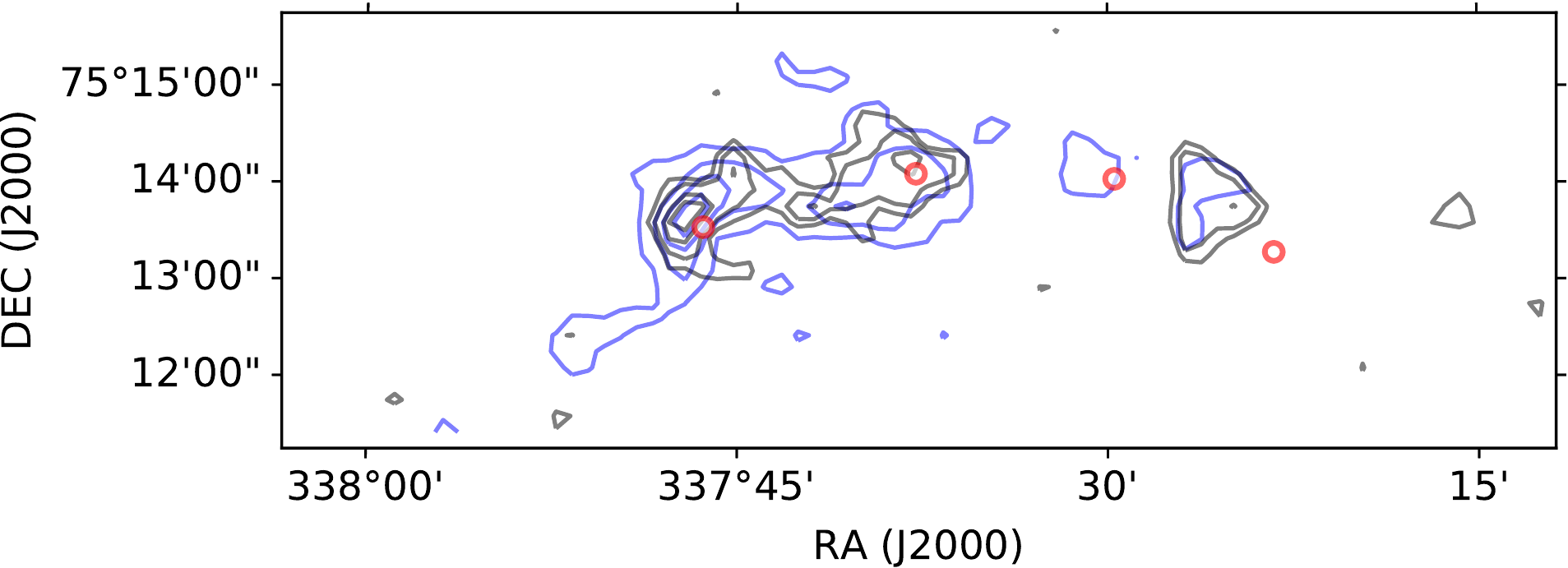}
\includegraphics[width=0.6\linewidth]{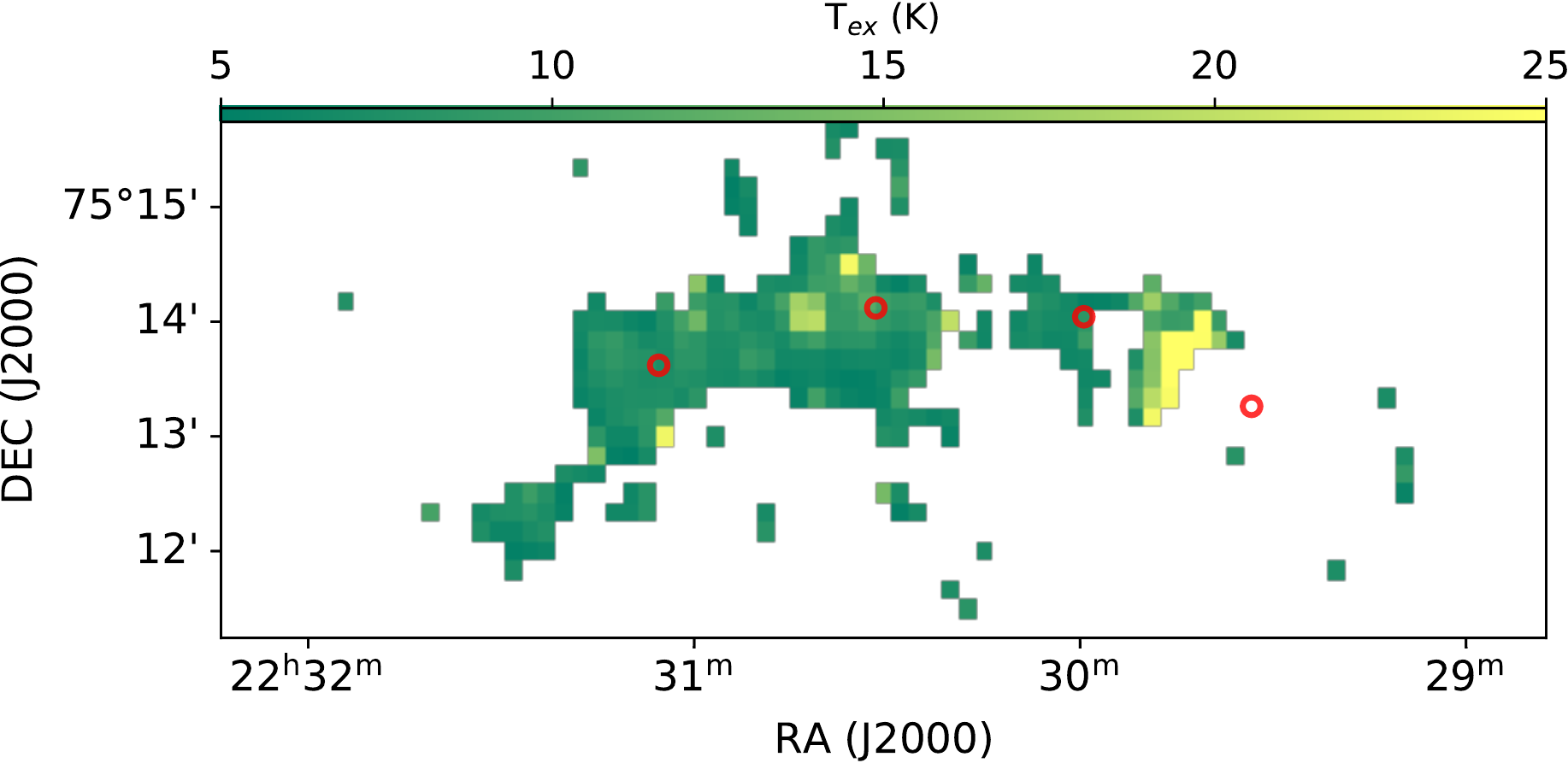}
\caption{   Upper panel: The blue contours show the integrated intensities of HC$_3$N $J$ = 10--9 observed using the NRO 45m, 
with levels from  0.1  to 0.35 and increments of 0.05 K km s$^{-1}$.
The black contours show the integrated intensities of HC$_3$N $J$ = 11--10 observed using the NRO 45m, with levels from 
0.06  to 0.18 and steps of 0.04 K km s$^{-1}$.
Lower panel: The excitation temperatures derived from  the ratios between the integrated intensities of  HC$_3$N $J$ = 11--10 and HC$_3$N $J$ = 10--9.
The four YSOs IRS1 to IRS4 are represented by red circles.
\label{fig:HC3N_45m}}
\end{figure*}

The   HC$_3$N $J$ = 11--10 and $J$ = 10--9 intensity maps detected by the NRO 45m, as well as the 
rotational temperature map derived from the ratios between the integrated intensities of these two lines 
are shown in Fig. \ref{fig:HC3N_45m}. The enhanced intensity ratios between HC$_3$N $J$ = 11--10 and 
$J$ = 10--9 near IRS1 may originate from the heating by IRS1. The rotational temperatures of HC$_3$N around 
IRS1 and IRS3 derived from the intensity ratios can be higher than 20 K (see the lower panel of Fig.
\ref{fig:HC3N_45m}). Because of the limited SNRs  of these two lines, the uncertainties of the derived rotational 
temperatures can be as high as 5 K. However, it can be confirmed that gas around these YSOs tends to be heated. 

The radiation from the YSOs may also play an important role in the evolution of the CCM abundances in the 
ambient matter around the protostellar sources. In star formation regions, HC$_3$N as well as HC$_5$N would 
be destroyed by stellar activities \citep{Taniguchi_2018}. From Fig. \ref{fig:HC3N_45m}, a clue for the 
erosion of HC$_3$N emission by IRS1 can be seen in the western margin of  the clump between IRS1 and IRS2. 
The enhancements of S-bearing CCMs induced by shocks may also be decomposed by stellar radiation. In 
evolved stage of SCCC, outflows become weaker while the effects of stellar radiation are  
getting stronger because of diminishing shielding effects from ambient matter around young stars. This may be the 
reason why SCCC sources with $X$[C$_3$S] exceeding $X$[HC$_5$N] are rare.

\section{Discussion} \label{sec:dis}
\subsection{Radial profile of F1-M}
For each of the extracted filaments F1-6 (see Sect. \ref{sec:fil_ext}), Gaussian fitting is applied to the 
radial surface density profile averaged along the main branch (see the right panel of Fig. \ref{fig_profile}).
The fitting to F1 (L1251-A) gives an FWHM radial width of  0.16$\pm$0.02 pc (1.8$\pm$0.2 arcmin), which is 
smaller than the values of other sub-filaments ($\sim$ 0.3 pc) in this region (see Table \ref{tab:fi_par}).
For F1 (L1251-A), a width  of 0.3 pc was derived by \citet{2016A&A...586A.126L} from NH$_3$ emission with an 
angular resolution $\sim$40\arcsec.   This difference can be explained by the higher angular resolutions 
of the Herschel data ($\sim$ 10\arcsec), which resolve  more centrally condensed structures near the ridges of 
filaments. From the right panel of Fig. \ref{fig_profile}, it is clear that single Gaussian component fits applied to the radial
profile of F1 are not as good as for other sub-filaments except for F4. If the radial  profile of F1 is fitted with 
two Gaussian components, the width of the broader one will be consistent with the widths of other substructures 
with a value $\sim$0.3 pc. We have also fitted the radial profile of F1 with a Plummer-like function 
\citep{1964ApJ...140.1056O,2016A&A...586A..27K}
\begin{equation}
\Sigma(r) = \Sigma_0/ \left[ 1+\left({r}/{r_0}\right)^2 \right]^{ (\alpha-1)/2},
\end{equation} 
assuming F1 is cylinder-like and oriented parallel to the plane of the sky. The fitted results are shown as red 
dashed lines in the right panel of Fig. \ref{fig_profile}. r$_0$ and $\alpha$ are obtained to be 0.05 pc and 2.5, 
respectively. The corresponding volume density of H$_2$   and linear density 
($\mathcal{M}_l$) in the central regions are $3\times 10^{4}$ cm$^{-3}$ and $\sim$36 M$_\sun$ pc$^{-1}$, respectively.

The Plummer distribution with $\alpha=4$ describes an 
isothermal gas cylinder in hydrostatic equilibrium. 
It gives $n(r)\propto r^{-4}$ when $r\to\infty$.
If $\alpha<2$, the linear density contributed from  the 
mass enclosed in the cylinder with radius of r ($\mathcal{M}_l(r)$) is infinite when $r\to\infty$. Thus a
cutoff of radial profile and nonlocal support mechanisms against gravity such as turbulence, magnetism and 
rotation must be introduced to stabilize the filament. The effective velocity dispersion ($\sigma_{\rm 1D}$) 
can be expressed as
\begin{equation}
\sigma_{\rm 1D}^2 = kT/\mu m_H+\sigma_{NT}^2+v_A^2/(3-D)+(r\Omega)^2/(3-D),
\end{equation}
where $v_A=\sqrt{\frac{B}{4\pi \rho}}$ is the Alfv\'{e}n speed, $\sigma_{NT}$ the non-thermal velocity 
dispersion, $\Omega$ the rotational angular  velocity, and D is the dimension of the considered structures 
(0 for isotropic spheres, 1 for filaments and 2 for sheets). Virial equilibrium requires 
$\sigma_{\rm 1D}\propto r^{(1-\alpha/2)}$ 
for Plummer distributions with $\alpha<2$. For a Plummer distribution with $\alpha>2$, the linear density 
$\mathcal{M}_l$ is finite, and the requirement of virial equilibrium leads to 
\begin{equation}
\sigma_{\rm 1D}\to \sqrt{G\mathcal{M}_l/2}\ \ \mathrm{when}\  r\to\infty\ ,
\end{equation}  
where the confined pressure is ignored. For thermal pressure supported filaments, the above formula leads 
to an expression of the critical linear mass ($\mathcal{M}_{l,c}$), i.e. 
\begin{equation}
\mathcal{M}_{l,c} = \frac{2k\bar{T}}{G\mu m_H}
\end{equation}
where $\bar{T}$ is the  average temperature.

For F1 with $\alpha=2.5$ and $\mathcal{M}_l\sim 36$ M$_\sun$ pc$^{-1}$, the dynamical equilibrium requires
$\sigma_{\rm 1D}\sim$ 0.3 km s$^{-1}$ which is approximate two times of the sound  speed considering 
$T_{\rm dust}\sim$ 10 K for F1. F1 will be gravitationally unstable if only the thermal pressure is considered, 
and this is contrary to the result of \citet{2016A&A...586A.126L} since the linear density we derived is larger
than the value they derived from NH$_3$ lines. The reason may be that the  NH$_3$ lines mainly trace the dense 
subsonic region while the masses derived from the dust continuum are dominated by the supersonic envelopes. The 
linear mass and Plummer index ($\alpha$) of F1 are similar to the values (41 M$_\sun$ pc$^{-1}$ and 2.2) for the 
Serpens filament, which is at the onset of a slightly supercritical collapse \citep{2018A&A...620A..62G}. The
$\sigma_{\rm 1D}$ required to stabilize F1 is comparable to the observed velocity dispersion 
($\delta V= \Delta V/\sqrt{8 \log(2)}$). Overall, F1 is supercritical 
\citep[M$_l$$>$M$_{l,c}$;][]{2019A&A...629L...4A} and mainly supported by turbulence. The gradual radial 
collapse and fragmentations along the ridge tend to change the $\alpha$ from 4 (the value for an isothermal 
cylinder)  to 2 (the value for clumps supported by uniform $\sigma_{\rm 1D}$).

\subsection{Different filamentary components in L1251-A}
\begin{figure*}
\centering
\includegraphics[width=0.5\linewidth]{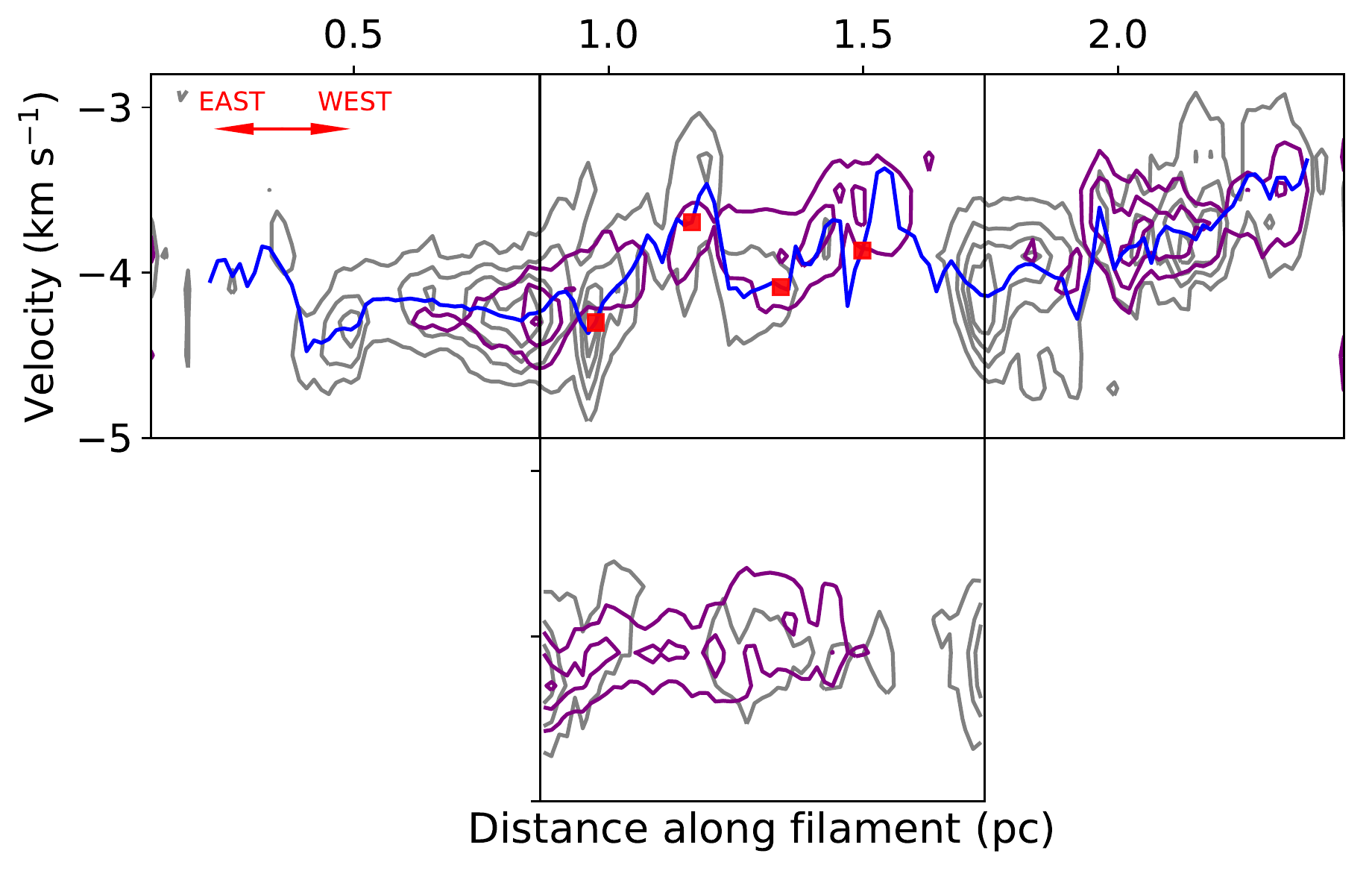}
\caption{ The P-V diagram of HCO$^+$ $J$ = 1--0 (gray contours) and SO $J_N$ = 2$_2$--1$_1$ 
(purple contours) along F1-M (upper panel) and F1-S (lower panel). 
Blue lines show the fitted central velocity of HCO$^+$ $J$ = 1--0, and
the red filled squares represent the locations of the four YSOs (IRS1 to IRS4 from right to left).
\label{fig:pv}
}
\end{figure*}

The position-velocity (P-V) diagrams of HCO$^+$ $J$ = 1--0 and SO 2$_2$--1$_1$ along the main branch of L1251-A 
(F1-M) and the side branch (F1-S) are shown in Fig. \ref{fig:pv}. The velocity patterns have no large scale 
gradient along F1-S, but show an oscillating pattern along F1-M with the material around IRS1 and IRS3 having 
obviously redder velocities (by 0.2-0.5 km s$^{-1}$) than those around the other two YSOs (IRS2 and IRS4) and 
those around F1-S. There is  a blue velocity lobe around the position of IRS4 and a red velocity lobe around IRS3 
(see Fig. \ref{fig:pv}). The velocity shifts of the two lobes may originate from outflows and are comparable with 
the velocity difference between IRS3 and IRS4. The large scale radial and  tangential velocity gradients  in the 
main part of L1251-A  \citep{2016A&A...586A.126L} may originate from the blend of the two sub-filaments.

\begin{figure*}
\centering
\includegraphics[width=0.8\linewidth]{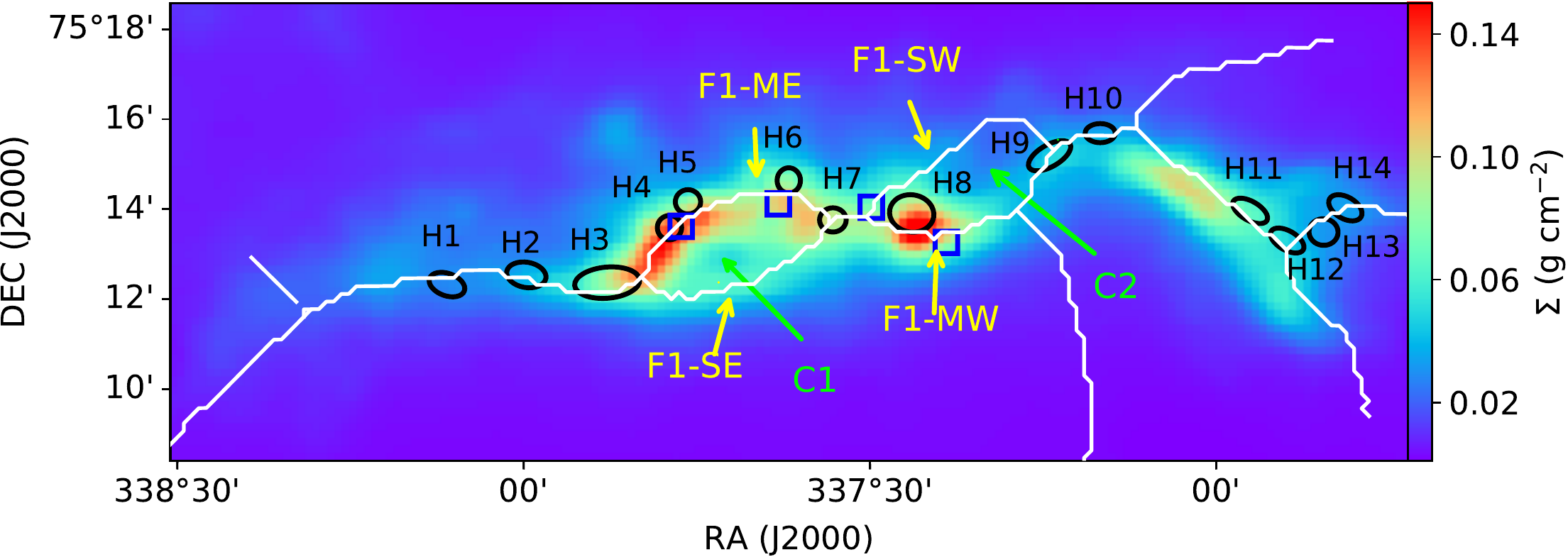}
\caption{The background shows the surface density of the dust in L1251-A derived from the  
dust continuum. 
The black ellipses represent the 
HCO$^+$ clumps identified using the Fellwalker algorithm (see Sect. \ref{sec_45m}). 
The four YSOs  are indicated as 
blue rectangles. White lines 
represent the filament skeleton. \label{fig:dust_hcopclumps} }
\end{figure*}

\begin{figure*}
\centering
\includegraphics[width=0.49\linewidth]{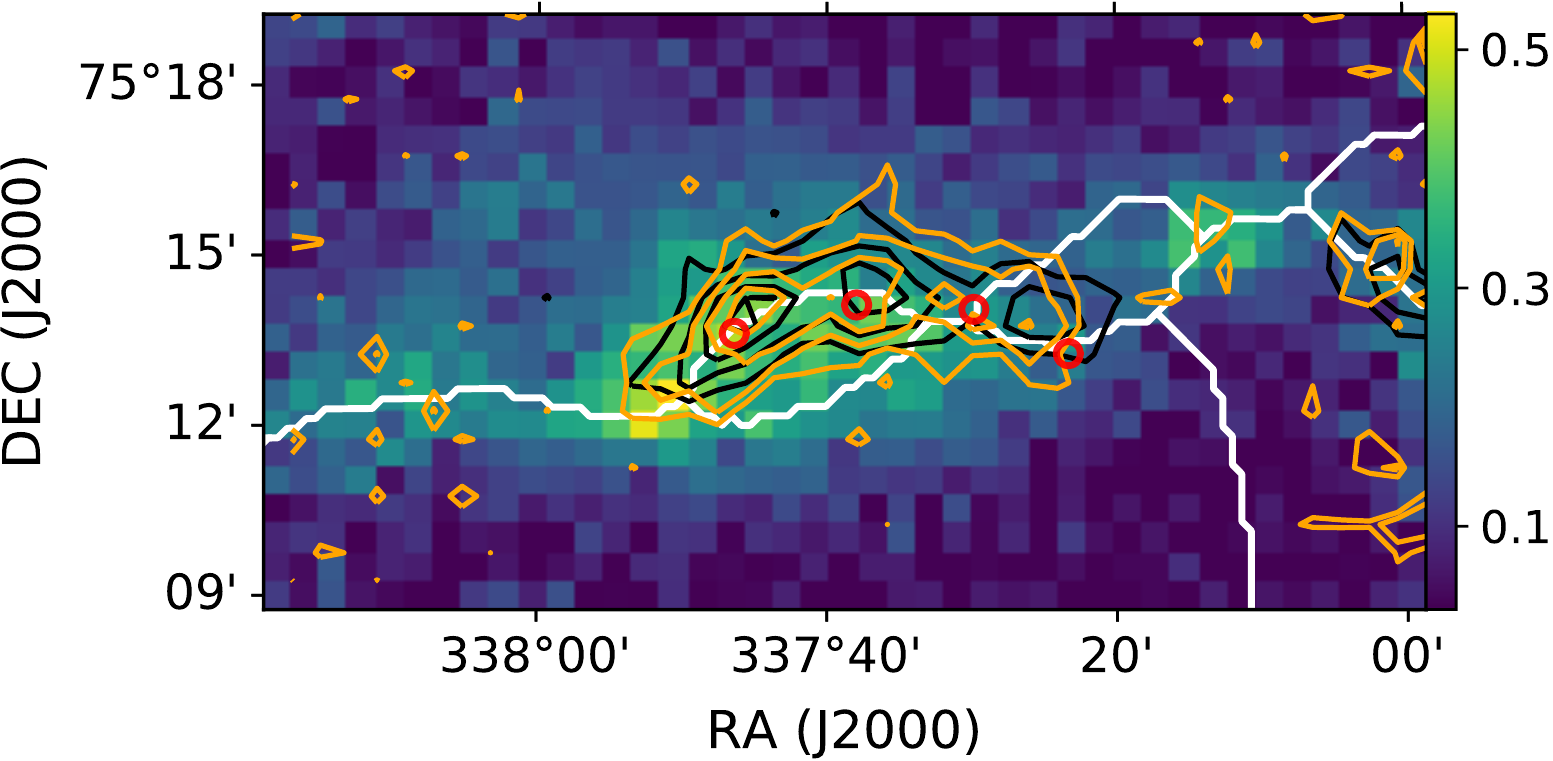}
\caption{The background, the orange contours and the black contours represent 
the integrated intensities (in units of K km s$^{-1}$) of CS $J$ = 2--1, 
C$_2$H $N$ = 2--1,  $J$ = 3/2--1/2, $F$ = 2--1 and N$_2$H$^+$ $J$ = 1--0, $F_1$ = 2--1,
respectively, mapped by the PMO 13.7 m.
The contours have levels from 30 percent to 90 percent and increments of 20 percent, relative
to the maximum intensities.
The white lines show the skeletons of the filaments extracted from Herschel maps 
(see Sect. \ref{sec:fil_ext} and Fig. \ref{fig:dust_hcopclumps}).
The four YSOs IRS1 to IRS4 are represented by red circles. \label{fig:pmo_c2hn2hp}
}
\end{figure*}

\begin{figure*}[htb]
\centering
\includegraphics[width=0.8\linewidth]{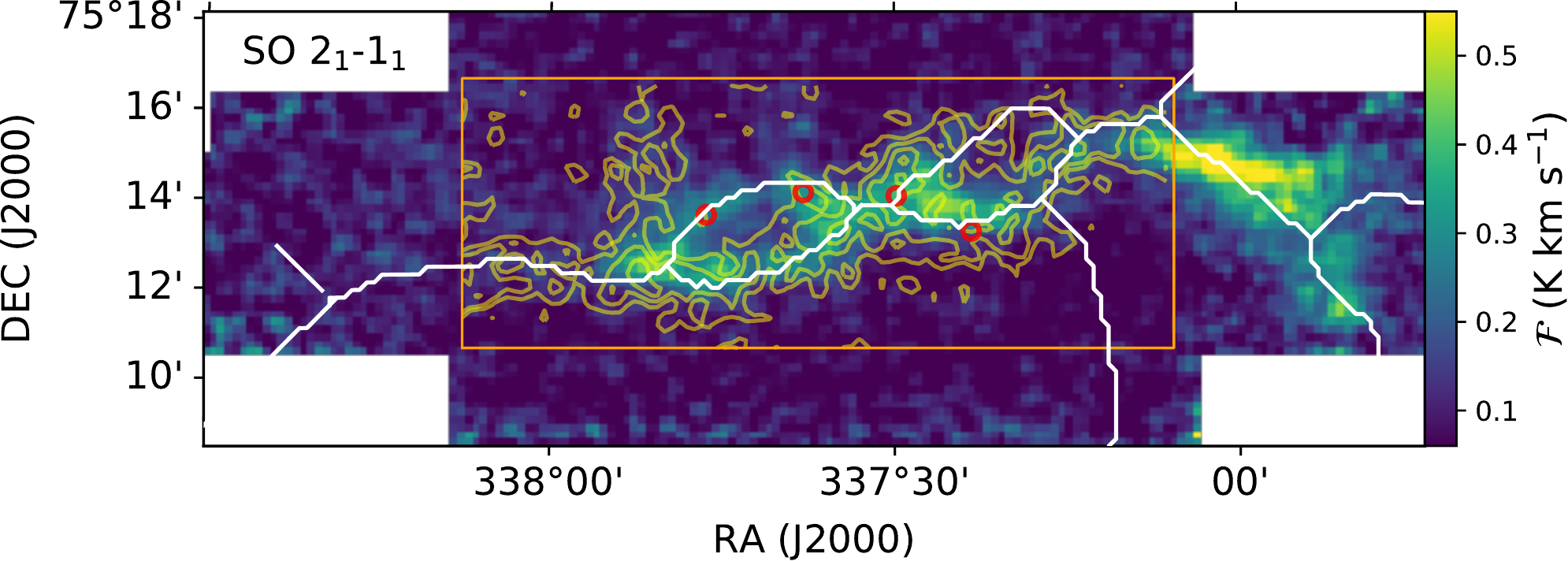}
\caption{Background is the integrated emission map of SO 2--1 line observed using the NRO 45m. Yellow contours
show the C$^{18}$O $J$ = 1--0 emissions   of the observed region enclosed in the orange box,
with levels from 0.5 to 1.1 and steps of 0.2 K km s$^{-1}$.  
The C$^{18}$O $J$ = 1--0 lines are integrated within the velocity range -4.7 km s$^{-1}$ to -3.5 km s$^{-1}$.
See Fig. \ref{fig:pmo_c2hn2hp} for the meanings of the white lines and red circles.
\label{map:int_SO}
}  
\end{figure*}

\begin{figure*}
\centering
\includegraphics[width=0.8\linewidth]{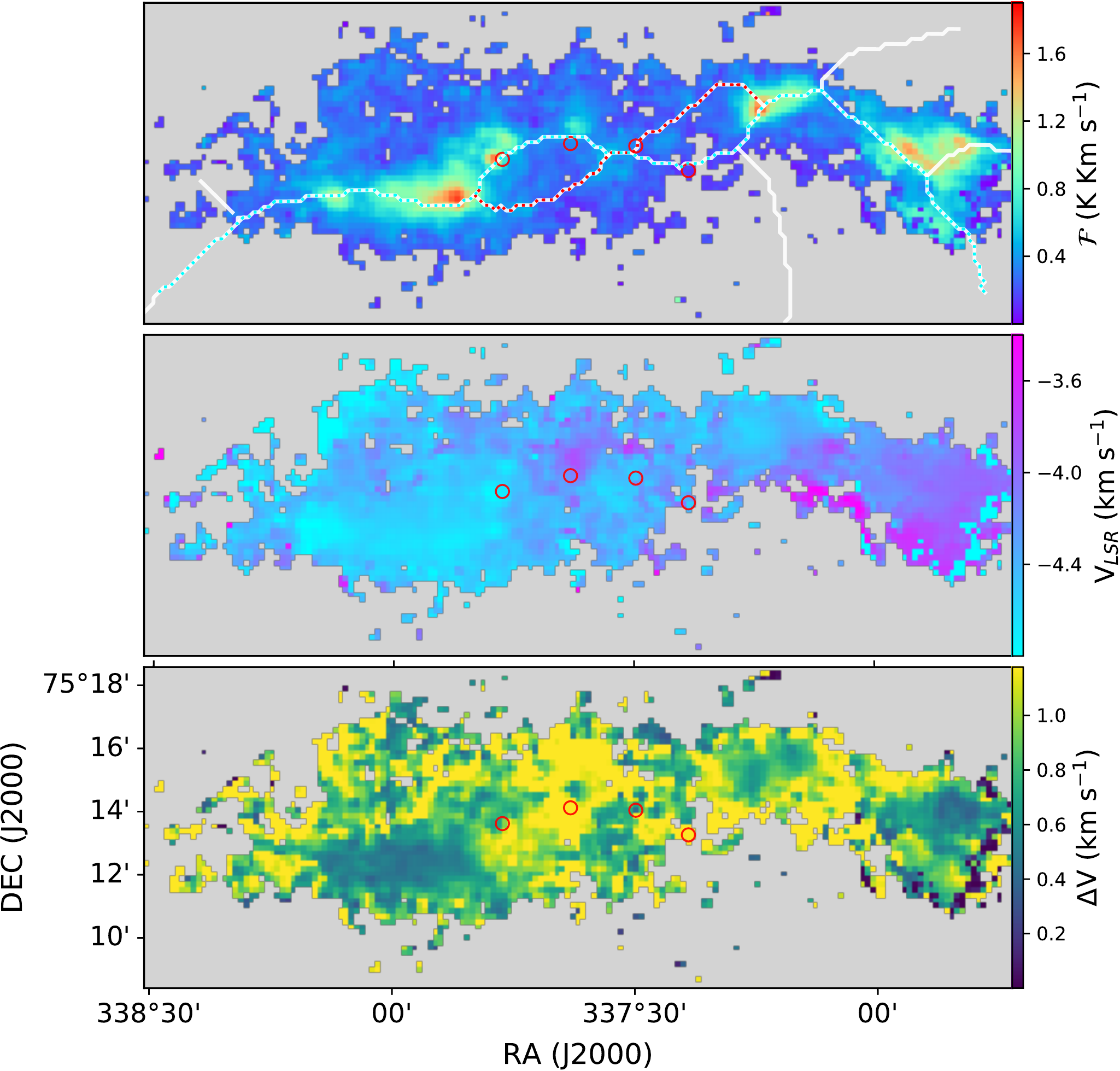}
\caption{The maps of integrated intensities (top),
$V_{\rm LSR}$ (middle) and line widths $\Delta V$ (bottom) 
from Gaussian fittings applied to HCO$^+$ $J$ = 1--0 observed using the NRO 45m. In the top panel,
white and cyan lines show the skeleton of F1. 
The cyan line represents the main branch of F1.
The dotted red line shows the side branch (F1-S) intertwined with the main branch.
The four YSOs IRS1 to IRS4 are represented by red circles. \label{fig:hcop_gpar}}
\end{figure*}

The length of F1-S is shorter than that of F1-M. 
The projection of F1-S on the sky plane reaching from the location of  H3, along
F1-SE and F1-SW, to the locations of H9/H10, and further spreading towards H11--H14 (see Fig. \ref{fig:dust_hcopclumps}). 
IRS1, IRS3 and IRS4 are 
located on F1-M, but it is not certain whether IRS2 is located on F1-M or F1-S. The F-M is more evolved than 
F-S, since the emission of dense gas traces such as N$_2$H$^+$ and HCO$^+$ tend to be stronger in F1-M, while
tracers of more diffuse gas such as CO are extensively distributed favoring F1-S. The line widths of HCO$^+$ 
around IRS2 are close to the values along F1-S, and lower than the values around the other three young stellar 
objects (see lower panel of Fig. \ref{fig:hcop_gpar}). Combining the morphologies of dust and molecular line 
emissions as well as the distribution of central velocities, we speculate that F1-S is closer to us along the 
line-of-sight with constant velocity, while F1-M is behind F1-S with velocities twisted by forming young 
stellar objects.

The surface density profile of the dust along the ridge of the main branch (from F3, F1, F5 to F6) is shown in Fig. 
\ref{fig_profile}. IRS3 and IRS4 coincide with their nearby local maxima of the dust profile with deviations smaller than 
20\arcsec~(0.03 pc). The other two YSOs (IRS1 and IRS2) is located at the right and left margins of a 
dust/HCO$^+$ clump (H8; see Fig. \ref{fig:dust_hcopclumps}) with distances from the center of H8 of about 
1\arcmin~(0.1 pc). The dust clumps and the dense gas clumps identified from HCO$^+$ emissions are not spatially 
coincident, but arrange in an alternate pattern (see Fig. \ref{fig:dust_hcopclumps}). The twisted spatial density and
velocity distribution along F1-M may be explained by a large-scale MHD-transverse wave  originating from 
outflows of young stellar objects \citep{2008ApJ...687..354N,2016A&A...590A...2S,2019MNRAS.487.1259L}. The 
transverse Alfv\'{e}n wave may also contribute to the assemblage of HCO$^+$ around the junction points of different 
filamentary branches. The gas components should couple well with the magnetic fields and flow sluggishly along the 
pipelines of magnetic fields. The dust is less affected, with peaks of the dust clumps misaligned with those 
of the dense gas clumps (see Figs. \ref{fig:dust_hcopclumps} and \ref{fig:hcop_gpar}). 
\citet{2019MNRAS.485.3991S}  investigated the non-linear evolution of the magnetized `resonant drag 
instabilities' (RDIs), and found the dust organizes into coherent structures and the system exhibits strong 
dust-gas separation. The separation between the gas and dust clumps can also be explained by instabilities 
induced by magnetic ambipolar diffusion along the filaments \citep{2018MNRAS.475.2632H,2018ApJ...864..108G}.

\subsection{Is C$_3$S a unique indicator of SCCC? } \label{abnornalc3s}
The abundance of sulfur  is quite uncertain. However, previous observations show that in cold 
cores and low mass star formation cores the column densities of C$_3$S are rarely higher than those of HC$_5$N. 
The C$_3$S column densities are even lower than those of HC$_7$N in all starless sources observed in  
Lupus I \citep{2019MNRAS.488..495W}. 

\begin{figure*}
\centering
\includegraphics[width=0.45\linewidth]{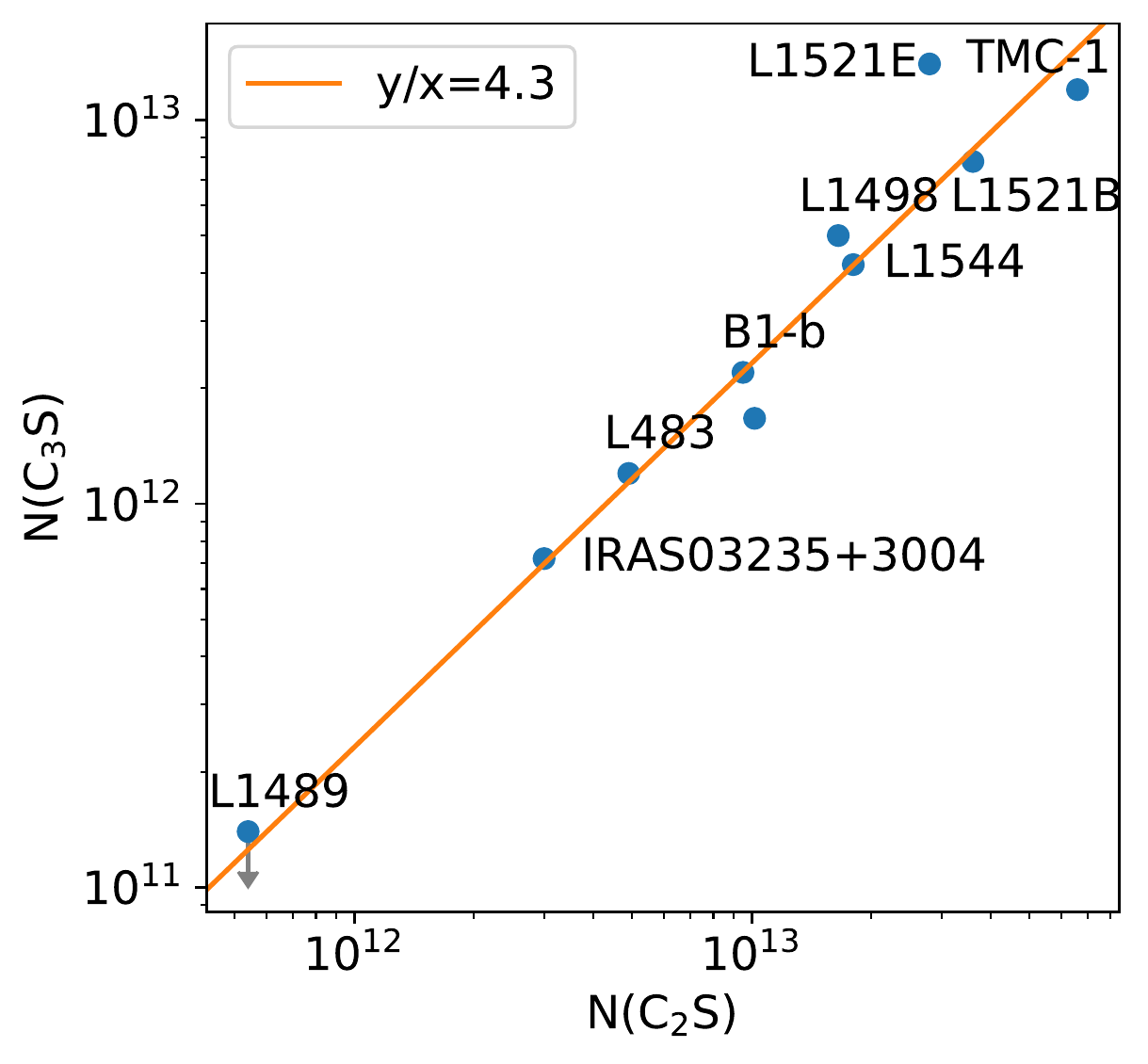}
\includegraphics[width=0.45\linewidth]{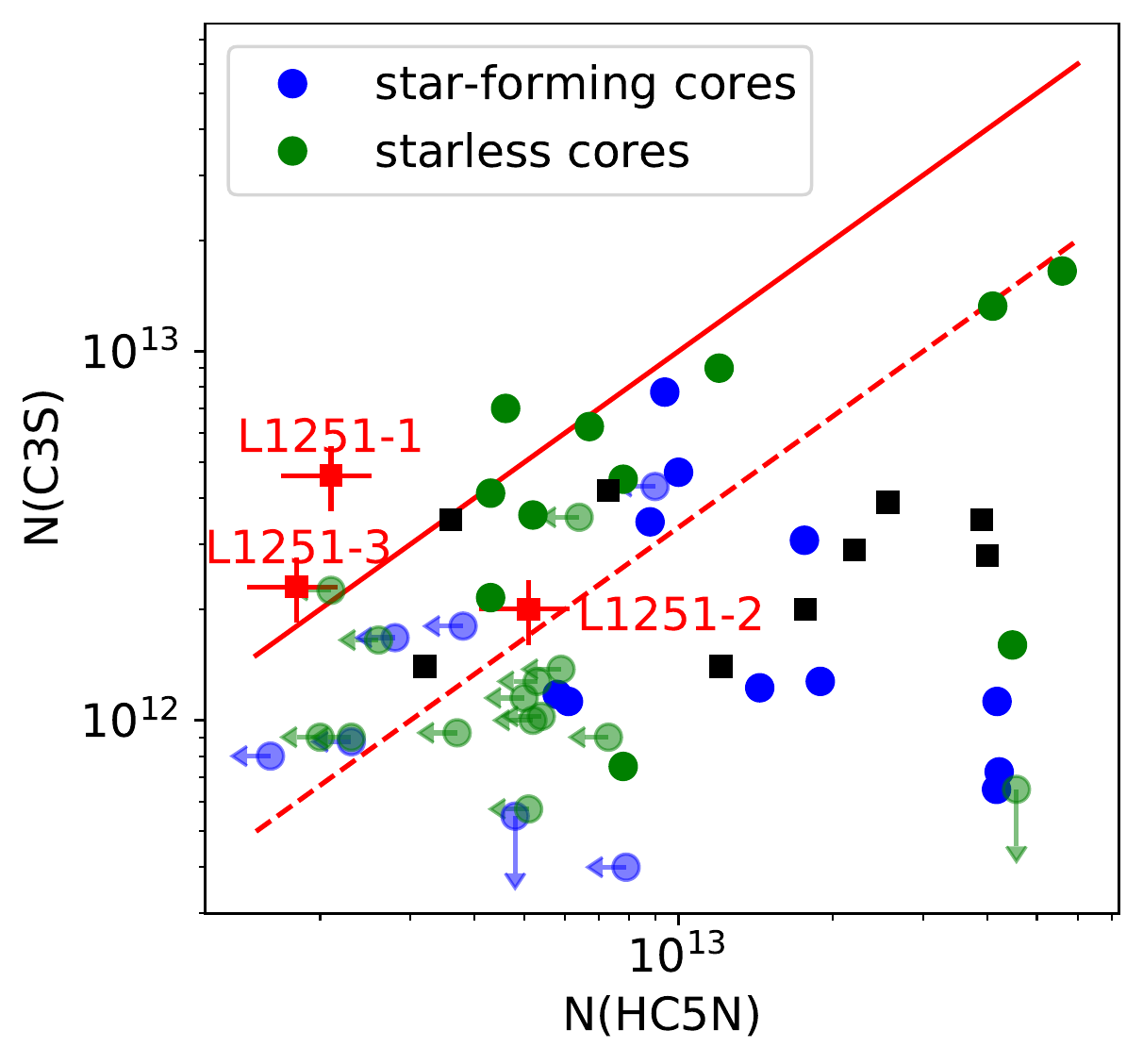}
\caption{ 
Left: the correlation between the column densities of C$_2$S and C$_3$S of low mass cores quoted from the 
literature \citep{2004ApJ...617..399H,2006ApJ...646..258H,2018ApJ...863...88L,2019A&A...625A.147A,2019A&A...630A.136N}.
Right: the correlation between the column densities of C$_3$S (derived from C$_2$S) and HC$_5$N quoted from \citet{1992ApJ...392..551S,2009ApJ...699..585H}.
The black filled squares represent sources detected in \citet{2019MNRAS.488..495W}.
The solid and dashed red lines represent $y=x$ and $y=x/3$, respectively. 
\label{figwhy}
}
\end{figure*}

The left panel of Fig. \ref{figwhy} shows the tight correlation between the column density of C$_2$S 
and C$_3$S in low mass cores quoted from the literatures. Using a conversion factor of 4.3 between $N$(C$_2$S) and 
$N$(C$_3$S), the column densities of C$_3$S can be obtained from those of C$_2$S since observations of low 
level transitions of C$_3$S are rarely reported.

The right panel shows the correlation between the column densities of C$_3$S (derived from C$_2$S) 
and HC$_5$N quoted from \citet{1992ApJ...392..551S,2009ApJ...699..585H}. All star forming sources have 
$N$(C$_3$S) lower than $N$(HC$_5$N). There is only one formerly studied starless sources L1521E having $N$(C$_3$S)$>$$N$(HC$_5$N). 
It seems impossible the cold gas components around young stars are responsible for the abnormally high abundance 
of C$_3$S  detected around YSOs in the L1251 region. Higher abundances of C$_3$S compared to HC$_5$N 
is a more strict criterion of SCCC in star-forming regions.

\section{Summary} \label{sec:summary}
The previously identified shocked carbon chain chemistry (SCCC) sources all have $N$(C$_3$S)$>$$N$(HC$_7$N) but 
$N$(C$_3$S)$<$$N$(HC$_5$N) deduced from Ku-band observations \citep{2019MNRAS.488..495W}. 
We thus searched for  SCCC sources, characterized by 
$N$(C$_3$S)$>$$N$(HC$_5$N), in radio K-band 
(18--26 GHz) using the Effelsberg 100m telescope. We identified L1251-1 as such a source, 
and L1251-3 as a candidate.
L1251-1 is located in L1251-A and harbors a Class Flat young 
stellar object (IRS1). 

We further mapped L1251-A at 3 mm band using the PMO 13.7 m 
telescope and the NRO 45 m telescope and in CO $J$ = 3--2 using the JCMT. Combining the 
obtained spectra and  archival data including Spitzer and Herschel continuum maps, we investigated the 
morphology, environment and dynamical characteristics of L1251-A as well as their relations to the phenomena 
of SCCC. The main results include:

\begin{itemize}
\item[1.] Molecular lines of NH$_3$, HC$_3$N, HC$_5$N, C$_3$S, c-C$_3$H$_2$ and C$_4$H were detected in three 
sources, L1251-1, L1251-2 and L1251-3. HFS and Gaussian fittings were applied to obtain basic parameters of the
detected lines. Column densities and abundances of these detected species were calculated.  

\item[2.] L1251-1 is characterized by $N$(C$_3$S) exceeding $N$(HC$_5$N) and is thus a confirmed SCCC source. 
L1251-3 has a  C$_3$S column density marginally larger than that of HC$_5$N. L1251-1 has similar abundance 
ratios of $N$(C$_4$H)/$N$(HC$_3$N) and $N$(HC$_3$N)/$N$(HC$_5$N) to the SCCC sources 
in \citet{2019MNRAS.488..495W}.

\item[3.] The volume densities of the three sources are constrained from the c-C$_3$H$_2$ 
1$_{1,0}$--1$_{0,1}$ emission line  and the   c-C$_3$H$_2$ 2$_{2,0}$--2$_{1,1}$ absorption line based on the 
RADEX model with an $ortho$-to-$para$ ratio of c-C$_3$H$_2$ assumed to be 3. We find the intensity ratio 
between c-C$_3$H$_2$ 2$_{2,0}$--2$_{1,1}$ and c-C$_3$H$_2$ 1$_{1,0}$--1$_{0,1}$ (noted as $\mathcal{R}$) can serve as 
a volume density tracer. This method gives a volume density of 3$\times$10$^5$ cm$^{-3}$ for L1251-2, and the 
values for L1251-1 and L1251-3 are estimated to be $\sim$ 10$^4$ cm$^{-3}$.

\item[4.] The L1251 region consists of hierarchical filaments in Herschel continuum maps. Six sub-filaments 
labeled as F1 to F6 were extracted using FilFinder. F1 associated with L1251-A has two intertwined 
branches, the main branch (F1-M) and the side branch (F1-S). F1-S is about 0.4 km s$^{-1}$ bluer than F1-M.
Four YSOs (IRS1, IRS2, IRS3 and IRS4) are likely located on F1-M, although there is also possibility that IRS2 
is on F1-S. 

\item[5.] The radial density distribution of F1 can be well fitted with a Plummer-like function. No large scale 
velocity gradients, neither radially nor  tangentially, are found in the main part of L1251-A.
 
\item[6.] IRS3 and IRS4 (L1251-2) both show strong jets and outflows. A possible shocked region  is found in 
the northwest of IRS4 and we name it IRS4-MNW. IRS4-MNW may be driven by the jet of IRS4. Outflow features are 
effective in this region and  we speculate that there were outflows driven by IRS1 although extinguished now.

\item[7.] The interstellar medium around IRS1 and IRS3 has redder velocities than those around the other two 
YSOs (IRS1 and IRS2) and those on F1-S. The twisted spatial density distribution and  velocity distribution 
along F1-M may present a large-scale MHD-transverse wave, resulting from outflows. 

\item[8.] The emission of dense gas tracers such as HCO$^+$ and N$_2$H$^+$ is associated with the main branch 
of F1 (F1-M). However, the emission of SO and C$^{18}$O is much more enhanced in the eastern part of the side 
branch of F1 (F1-SE) compared to F1-M. Principal component analysis is applied and confirms this characteristic.

\item[9.]  The peak positions of HCO$^+$ clumps are misaligned with those of the dust clumps.

\item[10.] SCCC in L1251-1 may have been caused by outflow activities from the infrared source IRS1. L1251-1 (IRS1)
together with the previously identified SCCC source L1251-IRS3 \citep{2019MNRAS.488..495W} 
demonstrate that L1251-A is an excellent region to study shocked carbon chain chemistry.
\end{itemize}

Overall, the outflow activities in L1251-A are responsible for  the physical characteristics including
the distorted velocity distribution along F1-M, the gas-dust misalignment, and the chemical property related to the 
abnormal high abundance of C$_3$S.

\begin{acknowledgements}
This project was supported by the National Key R\&D Program of China No. 2017YFA0402600, 
and the NSFC No. 12033005, 11988101, 11433008   
11373009, 11503035 and 11573036. 
D.Madones acknowledges support from CONICYT project Basal AFB-170002. Natalia Inostroza acknowledges CONICYT/PCI/REDI170243.

We wish to thank the staff of Effelsberg 100 m, NRO 45 m, PMO 13.7 m and JCMT 15 m for their support during the 
observations. The Effelsberg 100 m telescope is operated by the Max-Planck-Institut f\"{u}r Radioastronomie (MPIfR).
The Nobeyama Radio Observatory is a branch of the National Astronomical 
Observatory of Japan, National Institutes of Natural Sciences.
The PMO 13.7 m is operated by the Qinghai station of PMO at Delingha in China.
The James Clerk Maxwell Telescope is operated by the East Asian Observatory 
on behalf of The National Astronomical Observatory of Japan; Academia Sinica 
Institute of Astronomy and Astrophysics; the Korea Astronomy and 
Space Science Institute; Center for Astronomical Mega-Science. 
Additional funding support is provided by the Science and Technology Facilities 
Council of the United Kingdom and participating universities in the United Kingdom and Canada.

This research has made use of the NASA/IPAC Infrared Science Archive, 
which is operated by the Jet Propulsion Laboratory, 
California Institute of Technology, under
contract with the National Aeronautics and Space Administration.
\end{acknowledgements}

\software{
NOSTAR, 
FilFinder \citep{2015MNRAS.452.3435K}, 
PySpecKit \citep{2011ascl.soft09001G}, 
RADEX \citep{2007A&A...468..627V}, 
Starlink \citep{2014ASPC..485..391C}, 
GILDAS/CLASS 
}

\bibliographystyle{aasjournal}
\bibliography{draft}

\clearpage

\begin{table*}[htb]
\scriptsize{
\centering
\caption{The infrared and far-infrared continuum flux (in units of mJy) of YSOs in L1251. \label{tab:4stars}}
\begin{tabular}{lccccccccccccccc}
\hline
 YSOs$^1$ &  RA (J2000) & DEC (J2000) &    J & H & K$_s$ &   3.6$\mu$ &   4.5$\mu$ &   5.8$\mu$ &   8.0$\mu$ &   24$\mu$ &    70$\mu$ &   160$\mu$ &   350$\mu$ & 850$\mu$& Class\\
\hline
 IRS1 (L1251-1) &22:29:33.4 &+75:13:15.9&      -- &   -- &   -- &   5.6 &    10 &    15 &    18 &   55 &   150 &   302 &    --& --&Flat \\ 
 IRS2 & 22:29:59.5&+75:14:03.2 &   0.33 &  1.8 &  6.4 &    15 &    21 &    25 &    27 &  272 &   848 &  1532 &    --& 490&Flat\\ 
 IRS3  & 22:30:31.9 &+75:14:08.8&    -- &   -- &   -- &  0.13 &  0.42 &  0.32 &  0.16 &  5.0 &  1400 &  8241 &  9400& 1130&0/I\\ 
 IRS4 (L1251-2) & 22:31:05.6& +75:13:37.1&     -- &   -- &   -- &  0.43 &   1.1 &  0.75 &  0.37 &  1.9 &   688 &  3310 &  7500& 1570&II\\ 
L1251-3 &  22:37:31.13 &+75:10:41.5 &  -- &  -- &   -- &  0.06 &  0.08 & 0.035 & 0.248 & 0.82 &  30.1 &  182.1&    --& 1500& II\\
\hline
\end{tabular}\\
$^1${The data from the J (1.25 $\mu$m), H (1.65 $\mu$m),  K$_s$ (2.16 $\mu$m) bands (Cols. 4--6) 
are taken from the 2MASS All-Sky Point
Source Catalog (PSC) \citep{2006AJ....131.1163S}, 
and the magnitudes are converted into fluxes using the zero-magnitude attributes described in \citet{2003AJ....126.1090C}.
The 3.6 $\mu$m, 4.5 $\mu$m, 5.8 $\mu$m, 8.0 $\mu$m, 24 $\mu$m and 70 $\mu$m data (Cols. 7--12) 
were measured using the instruments IRAC and MIPs on Spitzer space telescope
by the c2d team \citep{2009ApJS..181..321E,2009ApJ...692..973E}.
The 160 $\mu$m data (Col. 13) are quoted from the PACS Point Source Catalogue in Herschel.
The 350 $\mu$m and 850 $\mu$m data (Cols 14--15) are obtained from \citet{2007AJ....133.1560W} and 
\citet{2017MNRAS.464.4255P}.
}
}
\end{table*}

\begin{table*}[htb]
\footnotesize{
\caption{Observed line and telescope parameters \label{table_lines_teles}}
\begin{tabular}{lccccccccc}
\hline
Molecules &	Transitions &	Freq. &	log$_{10}$(A$_{ij}$) &	S$_{ij}$ &	E$_u$ &	g$_u$ & 
HPBW$^1$  &	$\Delta_{ch}$          &	rms$^2$\\
        &       &   GHz   &                      &           & K      &       &  \arcsec     & km $s^{-1}$ &  mK\\
\hline
HC$_3$N &	$J$ = 2--1, $F$ = 3--2 &	18.196310 &	-6.88533 &	0.933 &	1.31 &	7 & 38$^E$ &0.63 &	10--60\\
HC$_3$N &	$J$ = 2--1, $F$ = 2--1$^3$ &	18.196217 &	-7.01030 &	0.5 &	1.31 &	5 & 38$^E$  &	0.63 &	\\
HC$_5$N &	$J$ = 7--6 &	18.638611 &	-6.18100 &	6.998 &	3.58 &	15 & 38$^E$	& 0.61 &	10--20\\
HC$_5$N &	$J$ = 8--7 &	21.301257 &	-6.00337 &	7.997 &	4.60 &	17 & 36$^E$	& 0.54 &	10--20\\
HC$_5$N &	$J$ = 9--8 &	23.963897 &	-5.84705 &	8.997 &	5.75 &	19 & 35$^E$	& 0.48 &	10--20\\
C$_3$S &	$J$ = 4--3 &	23.122983 &	-6.08154 &	4 &	2.77 &	9 &  35$^E$	& 0.49 &	10--20\\
NH$_3$  &	(1,1) &	23.694496 &	-6.77534 &	3 &	23.3 &	10 & 35$^E$	& 0.48 &	10--20\\
NH$_3$  &	(2,2) &	23.722633 &	-6.64906 &	6.664 &	64.4 &	6 & 35$^E$ &	0.48 &	10--20\\
C$_4$H  &	$N$ = 2--1, $J$ = 5/2-3/2, $F$ = 3--2 &	19.015143 &	-7.58621 &	2.8 &	1.37 &	7 &38$^E$ &	0.6 &	10--20\\
c-C$_3$H$_2$ &	1$_{1,0}$--1$_{0,1}$ &	18.343143 &	-6.37404 &	4.50 &	3.23 &	9 & 38$^E$ &	0.62 &	10--20\\
c-C$_3$H$_2$ &	2$_{2,0}$--2$_{1,1}$ &	21.587401 &	-6.20187 &	2.28 &	9.71 &	5 &	36$^E$ &0.53 &	10--20\\
HCO$^+$& $J$ = 1--0                  &   89.188523 & -4.38103 &  1   &  4.28    & 3   &   19$^{N}$ &  0.2 &  60 \\
SO      & 2$_2$--1$_1$            &   86.093950 & -5.27128 &  1.5 & 19.3     & 5   &  19$^N$    & 0.2  & 50 \\
HC$_3$N & $J$ = 10--9                 &   90.979023 & -4.23748 & 10   & 24.0     & 63  & 19$^N$     & 0.2  & 60 \\
HC$_3$N & $J$ = 11--10                &   100.07639 & -4.11136 & 11   & 28.8     & 69  & 17$^N$     & 0.2  & 50\\
C$_2$H  & $N$ = 1--0, $J$ = 3/2--1/2, $F$ = 2--1 &   87.316925 & -5.65605 & 2.4  & 4.2      & 5  & 19$^N$      & 0.2  & 110\\ 
        &                        &             &          &      &          &    & 56$^P$      & 0.21 & 55 \\ 
N$_2$H$^+$& $J$ = 1--0 $F_1$ = 2--1     &   93.173700 & -4.40926 & 5.37 & 4.47     & 15 & 19$^N$      & 0.2  & 170\\
        &                     &             &          &      &          &    & 56$^P$      & 0.2  & 74\\
C$^{18}$O & $J$ = 1--0                 &   109.78217 & -7.20302 & 1    & 5.27     & 3  & 15$^N$      & 0.2  & 150\\
CS      & $J$ = 2--1                 &   97.980950 & -4.77228 & 2    & 7.05     & 5  & 56$^P$      & 0.19 & 74 \\ 
CO & $J$ = 3--2                     &   345.79598  & -7.20302 & 3    & 31     & 7  & 15$^J$      & 0.2  & 350\\
\hline
\end{tabular} \\
The parameters of listed lines are adopted from splatalogue 
(\url{http://www.cv.nrao.edu/php/splat/advanced.php}).\\
$^1${The superscripts `E', `N', `P', and `J' in column 8 refer to the used facility including 
the Effelsberg 100 m, NRO 45 m, PMO 13.7 m and JCMT 15 m telescope, respectively.}\\
$^2${This column shows the 1-sigma rms noise levels on a brightness temperature scale.}\\
$^3${HC$_3$N $J$ = 2--1, $F$ = 2--1 and $F$ = 3--2 remain spectrally unresolved in our observations. }
}
\end{table*}

\begin{table*}
\caption{The parameters of the radio K-band lines except ammonia inversion transitions.  \label{tab:Kband_fit}}
\begin{tabular}{lcccccccc}
\hline
Transitions       &  HC$_3$N &  HC$_5$N      &  HC$_5$N     &  HC$_5$N    & C$_3$S     &     C$_4$H   & c-C$_3$H$_2$    & c-C$_3$H$_2$ \\
             &   $J$=2--1 $F$=3--2  &   $J$=7--6      &  $J$=8--7      &  $J$=9--8     &  $J$=4--3     &   $2_{\frac{5}{2}3}-2_{\frac{3}{2}2}$  &  1$_{1,0}$--1$_{0,1}$       &  2$_{2,0}$--2$_{1,1}$   \\
\hline
             &     \multicolumn{8}{c}{Integrated Intensities (K km s$^{-1}$)}                       \\
\cline{2-9}
    L1251-1 &  0.36(1) &  0.05(1) &  0.09(2) &  0.13(1) &  0.27(4) &  0.14(2) &     0.71(2) &    -0.27(1) \\
    L1251-2 &  0.65(2) &  0.20(1) &  0.19(1) &  0.32(1) &  0.12(1) &  0.12(1) &     1.34(1) &    -0.30(1) \\
    L1251-3 &  0.22(1) &  0.07(1) &  0.03(1) &  0.11(1) &  0.13(1) &  0.09(1) &     0.48(1) &    -0.19(1) \\
\hline
             &     \multicolumn{8}{c}{Line widths (km s$^{-1}$)$^{(1)}$}                       \\
\cline{2-9}
 L1251-1 &      1.3(1) &      1.0(2) &      1.2(4) &      0.9(1) &      2.5(5) &  1.8(4) &       1.4(1) &      1.2(1) \\
 L1251-2 &      1.4(1) &      1.4(2) &      1.3(1) &      1.2(1) &      1.0(1) &  1.3(2) &       1.4(1) &      1.3(1) \\
 L1251-3 &      1.5(2) &      1.4(5) &      1.1(5) &      1.3(2) &      1.4(2) &  1.9(4) &       1.5(1) &      1.4(1) \\
\hline
             &     \multicolumn{8}{c}{$V_{LSR}$  (km s$^{-1}$)}                       \\
\cline{2-9}
 L1251-1 &     -4.31(4) &     -4.4(1) &     -4.1(2) &     -4.14(6) &     -3.7(2) &  -4.2(1) &    -4.28(2) &    -4.14(4) \\
 L1251-2 &     -4.78(2) &     -4.8(1) &     -4.7(1) &     -4.72(4) &     -4.7(1) &  -4.7(1) &    -4.78(1) &    -4.64(3) \\
 L1251-3 &     -4.19(7) &     -4.1(2) &     -4.2(3) &     -4.02(9) &     -3.9(1) &  -4.1(2) &    -4.12(2) &    -3.97(5) \\
\hline
\end{tabular}\\
$^{1}$ The values of the  line widths and $V_{LSR}$ are not as accurate as the integrated intensities, because of the low velocity resolution mode ($\sim$0.5 km s$^{-1}$) chosen for the observations. The errors of line widths and $V_{LSR}$ listed here 
are given by corresponding HFS and Gaussian fittings.  \\
$^{2}$ The line widths and $V_{LSR}$ of HC$_3$N are obtained from HFS fittings. The parameters of other lines are obtained from
Gaussian fittings. 
A number in parentheses indicates the uncertainty in the last digit, not accounting for
an estimated 10\% uncertainty in amplitude
calibration.
\end{table*}

\begin{table*}
\caption{Parameters derived from HFS fittings of NH$_3$(1,1)  and Gaussian fittings of NH$_3$(2,2). \label{tab:NH3_pars}}
\begin{tabular}{lcccccccc}
\hline
       source &  v$_{11}$$^{(1)}$  & $\Delta V_{11}$ & T$_{MB,11}$ & $\tau$$_{11}$  &  $I_{22}$$^{(2)}$  \\
              & km s$^{-1}$  & km s$^{-1}$    &   K          &            &   K km s$^{-1}$         \\ 
\hline 
L1251-1       &-4.15(1)   &       0.95(1)   &     1.43(3)  &       1.72(8) &     0.14(2)\\
L1251-2       &-4.64(1)   &       0.87(1)   &     3.71(2)  &       2.59(3) &     0.60(1)\\
L1251-3       &-3.95(1)   &       1.07(3)   &     0.61(2)  &        0.6(1) &     0.05(1)\\
\hline
\end{tabular}\\
$^{1}${The errors of the v$_{LSR}$ and $\Delta V$ are mainly contributed by the spectral resolution ($\Delta_{re}\sim$ 0.5 km s$^{-1}$). }\\
$^{2}${The integrated intensity.}
\end{table*}

\begin{table*}
\caption{The parameters derived from spectra of NH$_3$ and c--C$_3$H$_2$, as well as the dust continuum. \label{tab:derivedpar1}}
\begin{tabular}{lcccccc}
\hline
            &  $T_{\rm ex}$(NH$_3$)$^{(1)}$ &  $T_{\rm rot}$(NH$_3$) &  $T_{\rm dust}$   & $n^{NH_3}$(H$_2$) & $n^{c-C_3H_2}$(H$_2$) & $N^{dust}$(H$_2$) \\
            &  K                &  K                 &      K        & $10^4$ cm$^{-3}$   & $10^4$ cm$^{-3}$ & 10$^{22}$ cm$^{-2}$ \\
\hline
   L1251-1  &  5.0              & 9.6(4)             &   10.4(3)     & 1.2         &  $\sim 1$ & 1.8(5)\\
   L1251-2  &  7.9              & 10.3(2)            &    10.3(3)    &  4.1             & 30 & 2.6(8) \\
   L1251-3  &  4.4              & 9.7(8)                &     10.0(3)   & 0.8            & $\sim 1$  & 1.5(4)\\  
\hline
\end{tabular}\\
$^1${This should be interpreted as a lower limit of the excitation temperature, considering the unknown beam filling factor.}

\end{table*}
  
\begin{table*}
\setlength{\tabcolsep}{0.05in}
\caption{The column densities.\label{tab:derivedpar}}
\begin{tabular}{lcccccc}
\hline
  Species   &  NH$_3$  &   HC$_3$N &  HC$_5$N &  C$_3$S &   C$_4$H & c-C$_3$H$_2$ $^2$   \\
\hline
            &  \multicolumn{6}{c}{Column Densities$^1$ }\\
\cline{2-7}
            & 10$^{14}$ cm$^{-2}$       & 1$0^{13}$ cm$^{-2}$     & $10^{12}$ cm$^{-2}$  & 10$^{12}$ cm$^{-2}$ & 10$^{13}$ cm$^{-2}$ &  10$^{13}$ cm$^{-2}$    \\
\hline
    L1251-1 &   4.1(5) & 1.68(9) &  2.1(3) &  4.6(7) &   11(2) &      $<$6.7  \\
    L1251-2 &   8.1(3) & 3.0(1) &    5.1(3) &  2.0(3) &    9(1) &     1.7  \\
    L1251-3 &   1.5(4) & 1.02(6) &  1.8(3) &  2.3(3) &    7(1) &      $<$4.8  \\
\hline
            &  \multicolumn{6}{c}{Abundances}\\
\cline{2-7}
            & 10$^{-8}$        & 1$0^{-10}$     & $10^{-10}$  & 10$^{-10}$ & 10$^{-9}$ &  10$^{-9}$    \\
\hline
    L1251-1 &   2.3(3) & 9.3(5)  &  1.2(2) &  2.6(4) &   5.6(9) &      $<$3.7 \\
    L1251-2 &   3.1(2) & 11.6(5) &  2.0(1) &  0.8(1) &   3.5(4) &      0.7  \\
    L1251-3 &   1.0(3) & 6.8(4)  &  1.2(2) &  1.5(2) &   4.0(7) &      $<$3.2 \\
\hline
\end{tabular}\\
$^1${The uncertainties of column densities are derived from errors of the Gaussian
fittings. The uncertainties of the abundances introduced by the uncertainties of $N$(H$_2$) 
and excitation temperatures 
are not included. }\\
$^2${The column densities and abundances of c-C$_3$H$_2$ for L1251-1 and L1251-3 
should be taken as upper limits (Sect. \ref{sec_c3h2}).}\\
\end{table*}

\begin{table*}
\caption{The basic parameters of the six sub-filaments. \label{tab:fi_par}}
\begin{tabular}{lccccc}
\hline
   & $L$$^1$ & $\Delta$ $^2$ & $L$/$\Delta$ &$\mathcal{M}_l$  & M  \\
   & pc  & pc &  & M$_\sun$ pc$^{-1}$ & M$_\sun$\\
\hline
F1 & 3.33(2) & 0.16(1) & 21(1)  & 36(1)  & 120(3)\\
F2 & 1.70(2) & 0.30(2) & 5.7(4) & 34(3)  &  58(6)\\
F3 & 2.29(2) & 0.31(1) & 7.4(3) & 45(2)  & 100(5)\\
F4 & 1.06(2) & --   & --  & --  & --\\
F5 & 2.38(2) & 0.28(1) & 8.5(3) & 3.0(1) & 7.0(3)\\
F6 & 4.46(2) & 0.28(1) & 11.0(5)  & 13.0(2)  & 58(1)\\
\hline
\end{tabular}\\
$^1${The length of the main branch. $^2${The fitted radial width.} }
\end{table*}

\begin{figure*}
\centering
\includegraphics[width=0.8\linewidth]{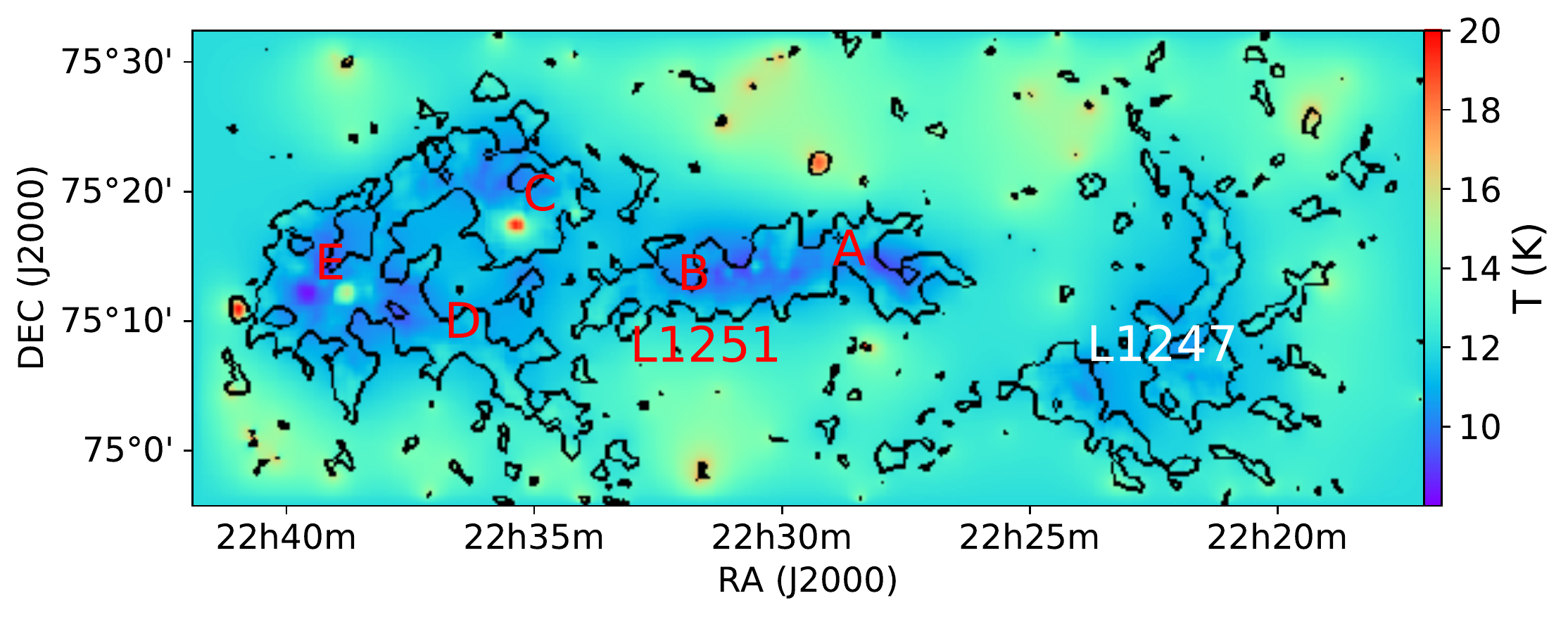}
\includegraphics[width=0.8\linewidth]{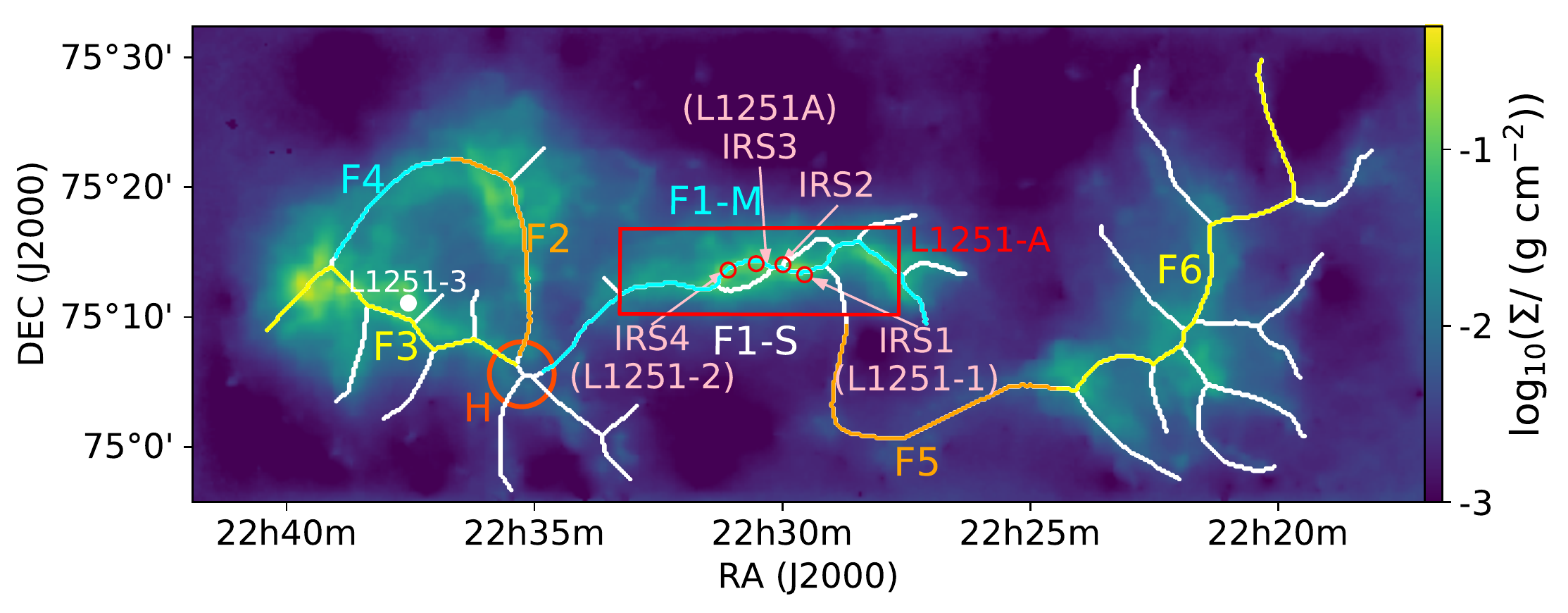}
\caption{Top: The dust temperatures  derived  from pixel-by-pixel greybody SED fittings using the  
160 $\mu$m, 250 $\mu$m, 350 $\mu$m and 500 $\mu$m continuum data from Herschel.
Only the pixels within the contours shown in the upper panel were fitted,  because the pixels 
outside have limited SNR in the 160$\mu$m band. The dust 
temperatures outside were extrapolated (see Appendix \ref{pixbypix_sed}).
Letters A through E label the positions of the five CO substructures 
of the L1251 region \citep{1994ApJ...435..279S}. 
Bottom: The surface density map of the dust. The red box represents our mapped 
region using PMO 13.7m and NRO 45m, 
and the four YSOs are represented by red open circles. IRS1 and IRS4 
are corresponding to L1251-1 and L1251-2. The white filled circle represents L1251-3. 
The white, cyan, orange, and yellow lines show the skeleton of filamentary structures 
extracted using FilFinder \citep{2015MNRAS.452.3435K}. 
The cyan, orange, and yellow colors are applied to the longest branches (main branches) 
of the six sub-filaments. 
The big vermilion circle shows the intersection region of F1, F2 and F3. 
F1--4 belong to L1251, and F6 is part of L1247.
\label{herscmap}}
\end{figure*}

\begin{figure*}
\centering
\includegraphics[width=0.4\linewidth]{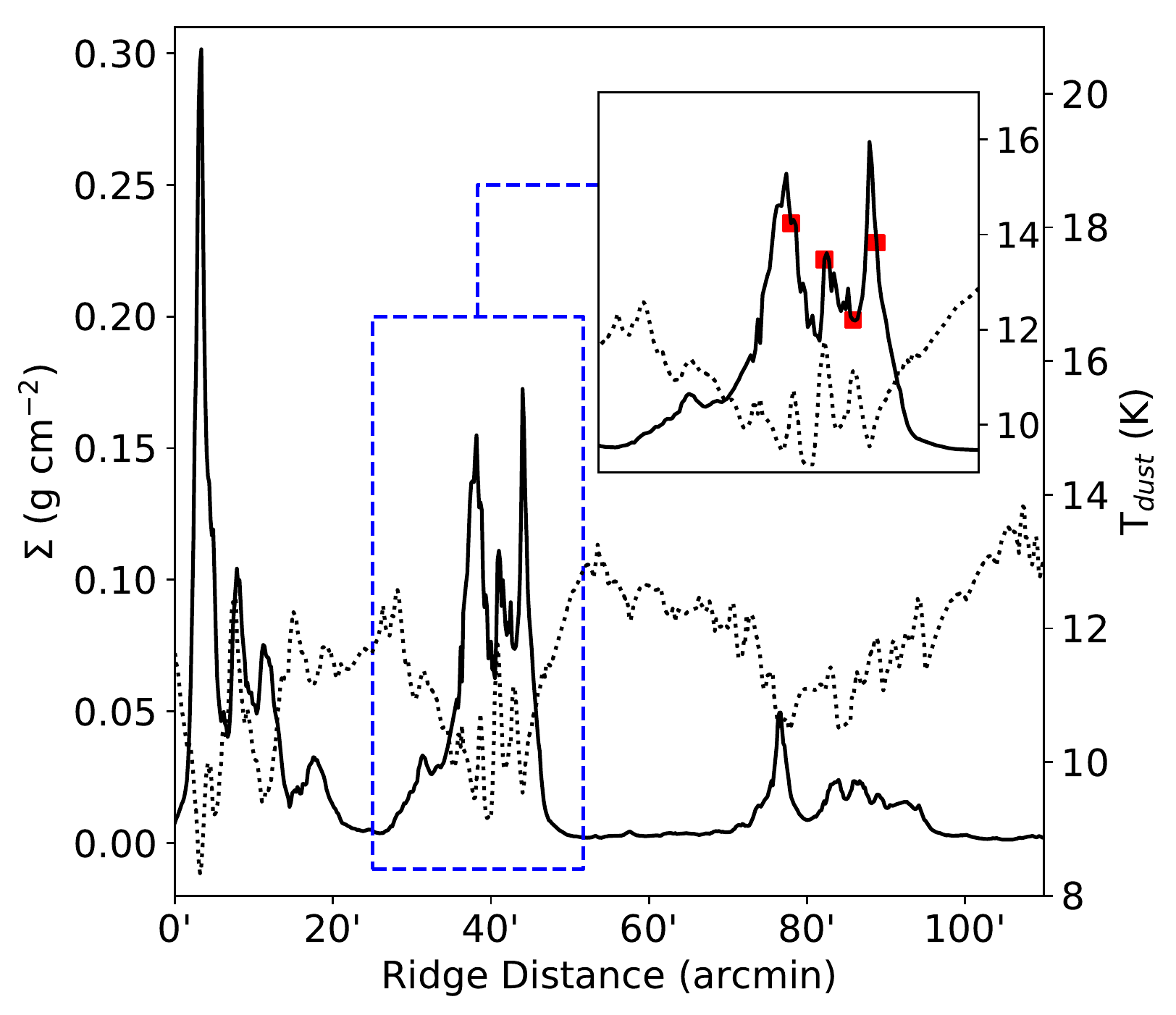}
\includegraphics[width=0.4\linewidth]{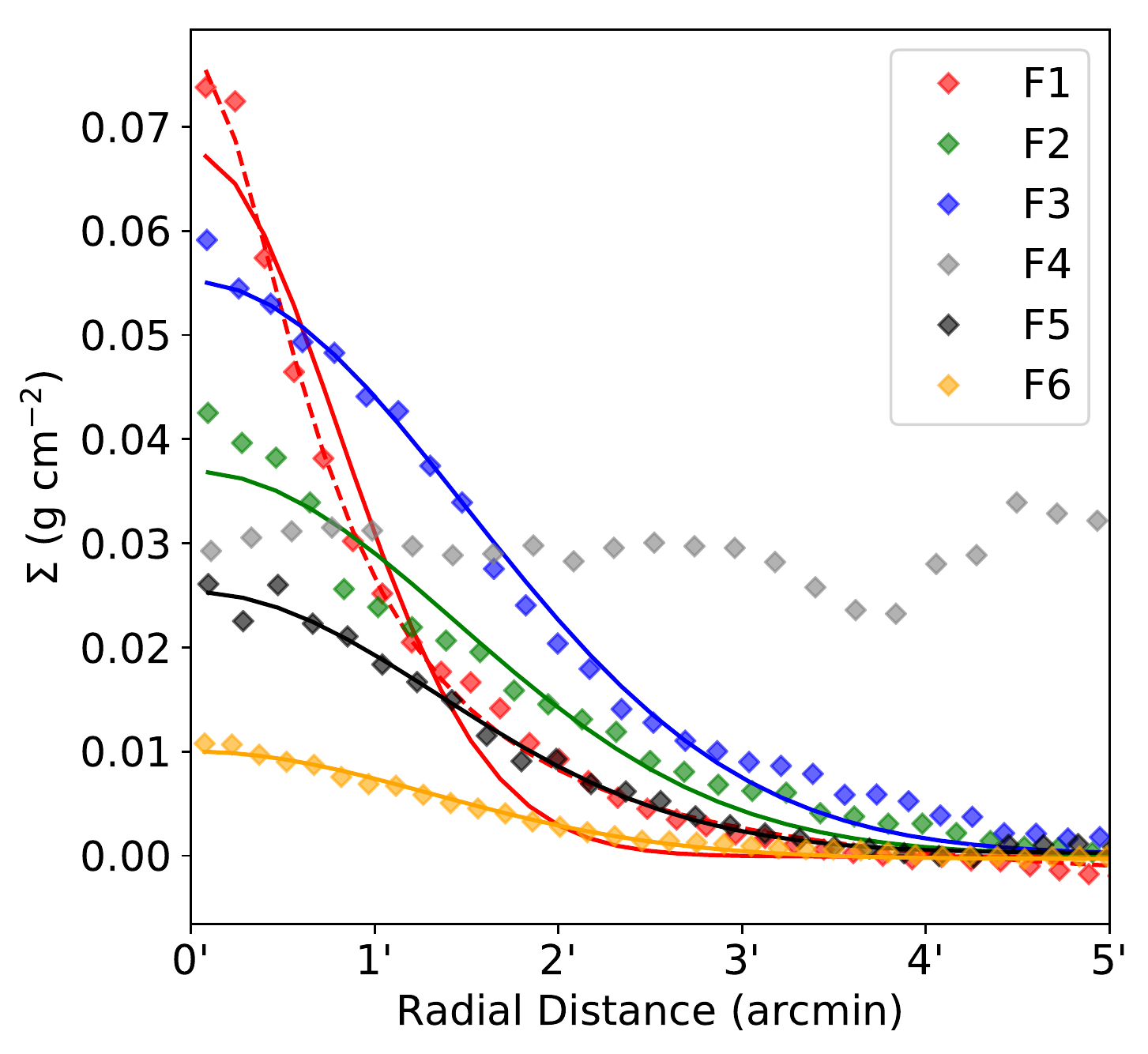}
\caption{Left: The profiles of surface density and dust temperature along the longest branch of the whole
filamentary structure shown in Fig. \ref{herscmap}, which is stitched from
the main branches of the sub-filaments F3, F1, F5 and F6.
The profiles enclosed in the dashed blue box, corresponding to F2, are zoomed in and shown at the 
top-right corner, with the red filled squares showing the locations of the four YSOs.
Right: The averaged radial surface density profiles of sub-filaments, with Gaussian fittings except for F4.
The surface densities of F5 have been multiplied by a factor of ten. 
\label{fig_profile}}
\end{figure*}

\clearpage
\appendix

\section{YSO Classfication}\label{sec_YSO}
YSO classfication based on the spectral index ($\alpha_\mathrm{IR}$) in near- and mid-infrared bands defined as $\alpha_\mathrm{IR}=d(\lambda L_\lambda)/d\lambda$
was first proposed by \citet{1987IAUS..115....1L}, and followed and improved by works such as 
\citet{1989ApJ...340..823W} and \citet{1994ApJ...434..614G}. 
The four class system for low mass YSOs based on $\alpha_\mathrm{IR}$ commonly used today can be described as
\begin{itemize}
\setlength{\itemsep}{2pt}
\setlength{\parsep}{2pt}
\setlength{\parskip}{2pt}
\item[]\textbf{I} $\alpha_\mathrm{IR} \ge 0.3 $
\item[]\textbf{Flat} $-0.3 \le \alpha_\mathrm{IR} < 0.3 $
\item[]\textbf{II} $-1.6 \le \alpha_\mathrm{IR} < -0.3 $
\item[]\textbf{III} $\alpha_\mathrm{IR} \le -1.6 $
\end{itemize}
\citet{1993ApJ...406..122A} suggested a class
``before'' Class I named Class 0, and the YSOs located between Class 0 and Class I
have protostellar masses  similar to that of  the remaining envelope.
IRS1, IRS2, IRS3 and IRS4 have spectral indices  of 0.19, 0.33, 0.70 and $-0.52$, 
respectively \citep{2009ApJS..181..321E}. 
\citet{2006ApJS..167..256R} found that YSOs with different spectral classes are distributed in different 
regions in the [3.6]-[5.8] v.s. [8.0]-[24] color-color diagram, and can also be used to classify
YSOs.  The four YSOs in the L1251-A region are all formerly located in the Class 0/I regions. 
IRS2 was re-classified  by \citet{2013AJ....145...94D} from Class I to Class Flat 
through de-reddening it using  an extinction value from the literatures.

We fitted the spectral energy distribution (SED) of these four YSOs using the model of  
\citet{2017A&A...600A..11R}. 
Since our fitted sources except IRS2 do not have data in 2MASS bands, the parameters of central stars and 
disks can not be well constrained.
The upper limits adopted as one tenth the values of IRS2 are fed to the fitting procedures, instead. The 
distances are fixed at 300 pc, 
the V-band extinctions (A$_V$) are limited in the range of 0--50 mag,
and the extinction law we adopted  is the same as that used in \citet{2010ApJ...716.1453F}. 
The fitted luminosities of central stars ($L_\star$) of IRS1, IRS2, IRS3 and IRS4 are 
0.3 $L_\sun$,1.4 $L_\sun$, 1.1 $L_\sun$ and 0.25 $L_\sun$,
respectively. If $A_V>40$ mag is permitted for IRS3 
\citep{2010ApJ...709L..74L}, corresponding to a deeply embedded young stellar object with 
a high accretion rate, the  unobscured $L_\star$ can be much higher than the value listed here.

All arguments considered, the IRS3 is a Class 0/I object, IRS4 is classified as Class II,
and  IRS1 and IRS2 are classified as Class Flat. 

\section{Greybody SED Fitting}\label{pixbypix_sed}
Pixel-by-pixel greybody SED fittings are applied to the Herschel archive data\footnote{\url{http://archives.esac.esa.int/hsa/whsa/}} 
covering the L1251 region in the 160,  250, 350 and 500 $\mu$m bands,
observed by \citet{2010A&A...518L.102A} as part of the Herschel Gould Belt Survey.  
The Herschel 70 $\mu$m data were not used in our fittings, because of limited SNR in this band.
We first filtered out the background emissions of each map following \citet{2017ApJS..231...11Y},
using the $CUPID-findback$ algorithm of the Starlink 
suite. 
Then, we convolve each Herschel map  
with a Gaussian Kernel with FWHM (in unit arcsec)
\[\sqrt{35.2^2-\theta_\lambda^2}\ ,\]
where $\theta_\lambda$ is the HPBW of the convolved Herschel map,
and 35.2 (in arcsec) is the HPBW of the Herschel 500 $\mu$m map.
The convolved maps were then re-gridded to obtain aligned images with spatial pixel size 
10\arcsec$\times$10\arcsec. For each pixel, its corresponding intensities at different wavelengths 
in the range of 160-500 $\mu$m are fitted with a blackbody model
\begin{equation}
I_\nu = B_\nu(T_{\rm dust})(1-e^{-\tau_\nu}),  \label{eq:Inu}
\end{equation}  
where  $B_\nu$ is the Planck function,  and the dust optical depth $\tau_\nu$
can be expressed as 
\begin{equation}
\tau_\nu = \Sigma\cdot\kappa_\nu/R_{gd}
\end{equation}
Here, $\Sigma$ is the surface density, and $R_{gd}$ is the gas-to-dust ratio which
is assumed to be 100. The dust opacity $\kappa_\nu$ can be expressed as a power
law in frequency,
\begin{equation}
\kappa_\nu = \kappa(600\ GHz)\left(\frac{\nu}{600\ GHz}\right)^\beta\ . \label{eq:kappa}
\end{equation}
The reference  dust opacity at 600 GHz (500 $\mu$m) can be adopted as 
5 cm$^2$ g$^{-1}$, the value for coagulated grains with thin ice mantles given by 
\citet{1994A&A...291..943O} (hereafter OH5).
In this work, $\kappa(600\ GHz)=$ 3.33 cm$^2$ g$^{-1}$ was adopted,
which is scaled down by a factor of 1.5 from the OH5 value as
suggested by \citet{2003A&A...399L..43B,2010ApJ...712.1137K}.
$\beta$ is fixed at 2. There are two free parameters, the dust temperature $T_{\rm dust}$
and the surface density $\Sigma$ to be obtained from the fitting procedures.

The uncertainty of the intensity at pixel (i,j) 
was taken as  $\sigma_{\nu;i,j} = \sqrt{(\sigma^{abs}_\nu)^2 + (\sigma^{rel}_{\nu;i,j})^2}$,
where  $\sigma^{abs}_\nu$ was estimated as the rms value of regions free from emission, 
while $\sigma^{rel}_{\nu;i,j}$ is taken as 15 percent of
the intensity of that pixel, based on the report by \citet{2013A&A...551A..98L}.
Only the pixels with intensities larger than $3\sigma_{\nu;i,j}$ at all four bands
were fitted by minimizing 
\begin{equation}
\chi^2_{i,j}=\sum_\nu (I_{\nu;ij}(T_{ij},\Sigma_{ij}) - I_{\nu;i,j}^{Herschel})^2/\sigma^2_{\nu;i,j}\ .
\end{equation}
The fitted maps of $T_{\rm dust}$ and $\Sigma$ are shown in Fig. \ref{herscmap}.

The fitted pixels are enclosed by the black contour shown in the upper panel of Fig. \ref{herscmap}.
The dust temperatures are as low as about 8 K in the central part of the fitted regions,
and increase to about 12 K at the margins. For pixels outside the fitted regions, the dust temperatures
are extrapolated by solving the Laplace's equation $\nabla^2 T=0$, with the values of the 
fitted regions unchanged and the values of the pixels at the margins of the whole map fixed as 12 K.
Numerically, $\nabla^2 T=0$ can be solved by iterating
\begin{equation}
T^{n+1}_{i,j} = \left\{
\begin{aligned}
& \frac{1}{4} \sum_{ 
 \Delta j \in \{-1,1\}   
} \sum_{\Delta i \in \{-1,1\}}   T^n_{i+\Delta i,j+\Delta j}
& for\ (i,j) \notin \mathcal{C} \\
&T^{n}_{i,j} & for\ (i,j) \in \mathcal{C}
\end{aligned}
\right.
\end{equation}
where $\mathcal{C}$ is the set consisting of the fitted pixels and the marginal pixels of the whole map.
The surface densities for pixels outside the fitted regions are calculated based
on the 500 $\mu$m intensities  through equation \ref{eq:Inu}
and \ref{eq:kappa}, and the extrapolated dust temperatures.

\section{Calculate Column Densities} \label{sec_nnh3}
The column densities of molecules  can be derived through 
\citep{2009ApJ...693.1736W,2015PASP..127..266M}
\begin{equation}\begin{aligned}
N=  \frac{3k}{8\pi^3\nu}\frac{Q_{\rm rot}\exp(E_u/kT_{\rm ex})}{S\mu^2} 
\frac{J(T_{\rm ex})}{J(T_{\rm ex}) - J(T_{\rm bg})} \int T_r d\upsilon  \label{eq:col_general}
\end{aligned}
\end{equation}
where h is Planck's constant, c is the speed of light, $k$ is the 
Boltzmann constant, S is the line strength, $\mu$ 
is the dipole moment, and $Q_{\rm rot}$ the partition function.

For NH$_3$(1,1), the optical depth ($\tau$), the intrinsic 
full-width at half power linewidth ($\Delta V_{\rm in}$), the LSR velocity 
($V_{\rm LSR}$) and the amplitude $\mathcal{A}$ can be obtained through
hyperfine structure (HFS) fitting 
$T_b(v) = \mathcal{A}(1-e^{-\tau(v)}) $ 
with
\begin{equation}
\tau(v)=\tau_{\rm tot}
\sum_0^{N-1}r_i\times exp\left(-4ln(2)\left(\frac{v-v_{\rm LSR}-v_i}
{\Delta V_{\rm in}}\right)^2\right) \label{tau_def}
\end{equation}
where where the $r_i$ are the relative
        intensities of the individual hyerfine (hf)
        lines in the optically thin case under
        conditions of thermodynamical equilibrium.
The excitation temperature ($T_{\rm ex}$) can be obtained from 
\begin{equation}
\mathcal{A}=\eta(J_\nu(T_{\rm ex})-J_{\nu}(T_{\rm bg})) \label{eq:exci}
\end{equation}
where $J_\nu(T)$ is defined by
$J_\nu(T)=\frac{h\nu/k}{e^{h\nu/kT}-1}$,
and $T_{\rm bg}$ is the background temperature.

For NH$_3$, the equation to calculate the column densities of molecules in the (1,1) and (2,2) states can 
be expressed as 
\begin{equation}
N_{JK}= \zeta_{JK} \mathcal A\tau_{\rm tot}\Delta V_{\rm in}
          \left(\frac{1}{\eta}+\frac{J_{\rm bg}+0.5h\nu/k}{\mathcal{A}}\right) \label{NH3_u_column}
\end{equation}
with  $\zeta_{11}=1.3850\times 10^{13}$, 
$\zeta_{22}=1.0375\times 10^{13}$ \citep{2013A&A...553A..58L}.
Rotational temperatures 
($T_{\rm rot}^{21}$) can also be derived from the population
ratio between the (2,2) and (1,1) states of NH$_3$.

The rotational temperature $T_{\rm rot}^{21}$ is adopted to calculate the total column density of NH$_3$ 
($N$(NH$_3$)) assuming only the metastable states are effectively populated,
\begin{equation}
N({\rm NH}_3) = N_{11} \sum_{i=0} \frac{g_{I;K=i}}{g_{I;K=1}}\frac{2i+1}{3} exp\left(\frac{-(E_i-23.2)}{T_{\rm rot}}\right)
\label{eq:Ntot}
\end{equation}
in which the ratio between the nuclear spin degeneracy ($g_I$) of $para$-NH$_3$ ($K\ne 3n$) and that of $ortho$-NH$_3$ ($K=3n$) is two.

The volume density can be estimated for T$_{\rm ex}\neq$T$_{\rm kin}$ through \citep{1983ARA&A..21..239H,2013A&A...553A..58L}
\begin{equation}
n({\rm H}_2) = \frac{A}{C}\left[\frac{J_\nu(T_{\rm ex}-J_\nu(T_{\rm bg}))}
{J_\nu(T_{\rm kin})-J_\nu(T_{\rm ex})}\right]\left[1+\frac{J_\nu(T_{\rm kin})}{h\nu/k}\right]
\end{equation} 
where $A$ is the Einstein A-coefficient and $C$ is the rate coefficient
for collisional de-excitation.

\section{Emission model of c-C$_3$H$_2$ } \label{c3h2_model}
We calculated the intensities of the c-C$_3$H$_2$ 1$_{1,0}$--1$_{0,1}$  
and c-C$_3$H$_2$ 2$_{2,0}$--2$_{1,1}$ lines produced 
in a uniform medium using the non-LTE radiative transfer code RADEX \citep{2007A&A...468..627V}.
The free parameters are the column density of $para$-C$_3$H$_2$ ($N^p$), 
the kinetic temperature ($T_{\rm kin}$), as well as the densities of the collision partners.
In our cases, H$_2$ is taken as the only collision partner for both  $ortho$-C$_3$H$_2$ and  $para$-C$_3$H$_2$, 
and the background emission is fixed as 2.73 K.
The space of the free input parameters ($n$(H$_2$), $N^p$, $T_{\rm kin}$) 
is gridded with $n$(H$_2$) and $N$(X) evenly sampled on a log scale stepped 
by a factor of three from 10$^3$ to 10$^7$ cm$^{-3}$ and 
from 10$^{11}$ to 10$^{15}$ cm$^{-2}$ respectively, and $T_{\rm kin}$ 
evenly sampled from 5 to 30 K stepped by 1 K.
For each set of input parameter, 
the integrated intensities (in units of K km s$^{-1}$) of c-C$_3$H$_2$ 1$_{1,0}$--1$_{0,1}$ ($\mathcal{F}^o$) and 
that of c-C$_3$H$_2$ 2$_{2,0}$--2$_{1,1}$ ($\mathcal{F}^p$) were 
calculated. Via extrapolation, the 3D surfaces $\mathcal{F}^p$($n({\rm H}_2)$, $N^p$, $T_{\rm kin}$) and 
$\mathcal{F}^o$($n({\rm H}_2)$, $N^o$, $T_{\rm kin}$) can be obtained.
Since the o/p ratios are fixed at three, the flux ratio between c-C$_3$H$_2$ 1$_{1,0}$--1$_{0,1}$ 
and 2$_{2,0}$--2$_{1,1}$ ($\mathcal{R}$) can be expressed as 
\begin{equation}
\mathcal{R}(n({\rm H}_2), N^p, T_{\rm kin}) = \frac{ \mathcal{F}^p(n({\rm H}_2), N^p, T_{\rm kin})  }{ \mathcal{F}^o(n({\rm H}_2), 3N^p, T_{\rm kin}) } \label{eq_R}
\end{equation}

\begin{figure}[htb]
\centering
\includegraphics[width=0.8\linewidth]{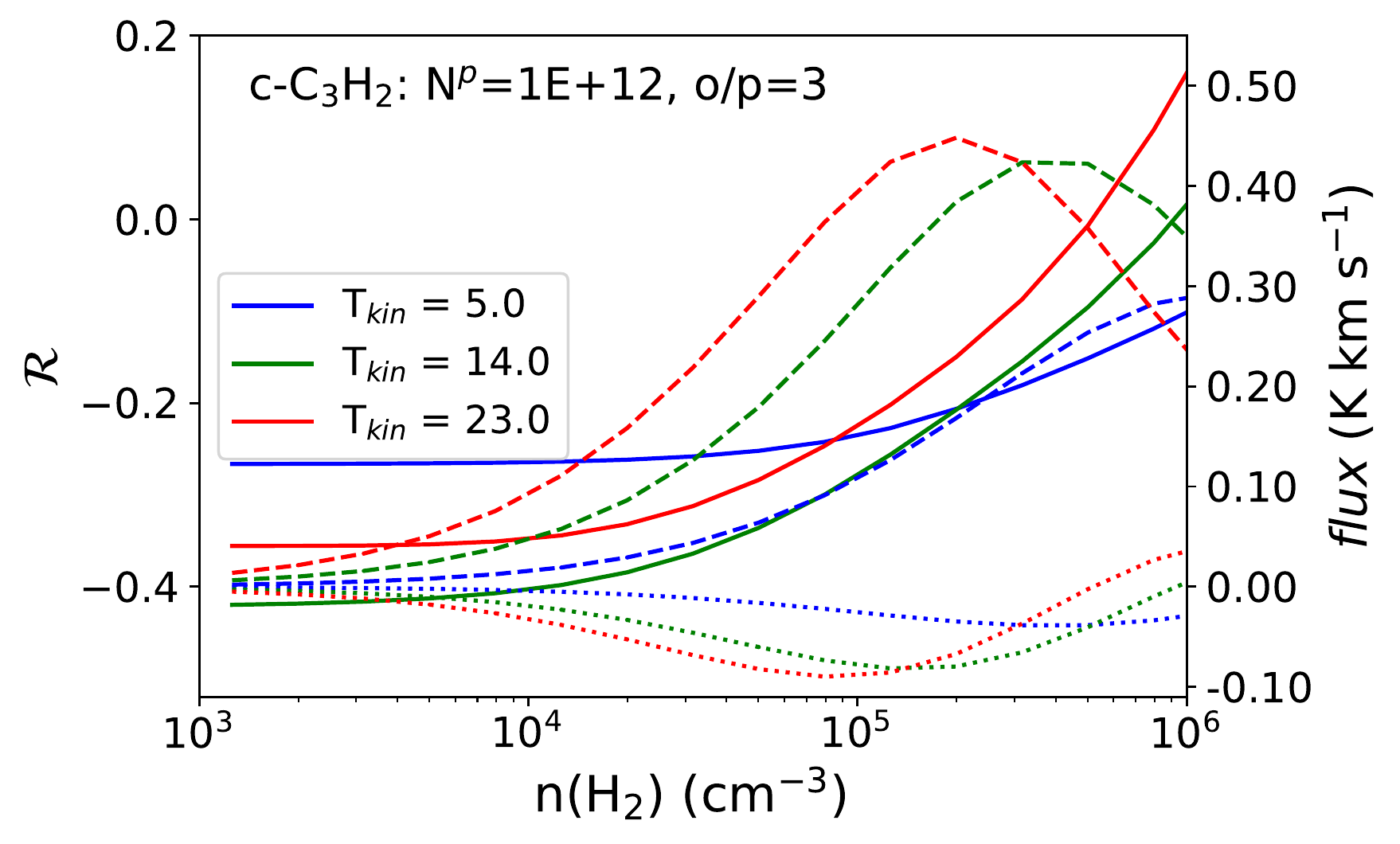}
\includegraphics[width=0.8\linewidth]{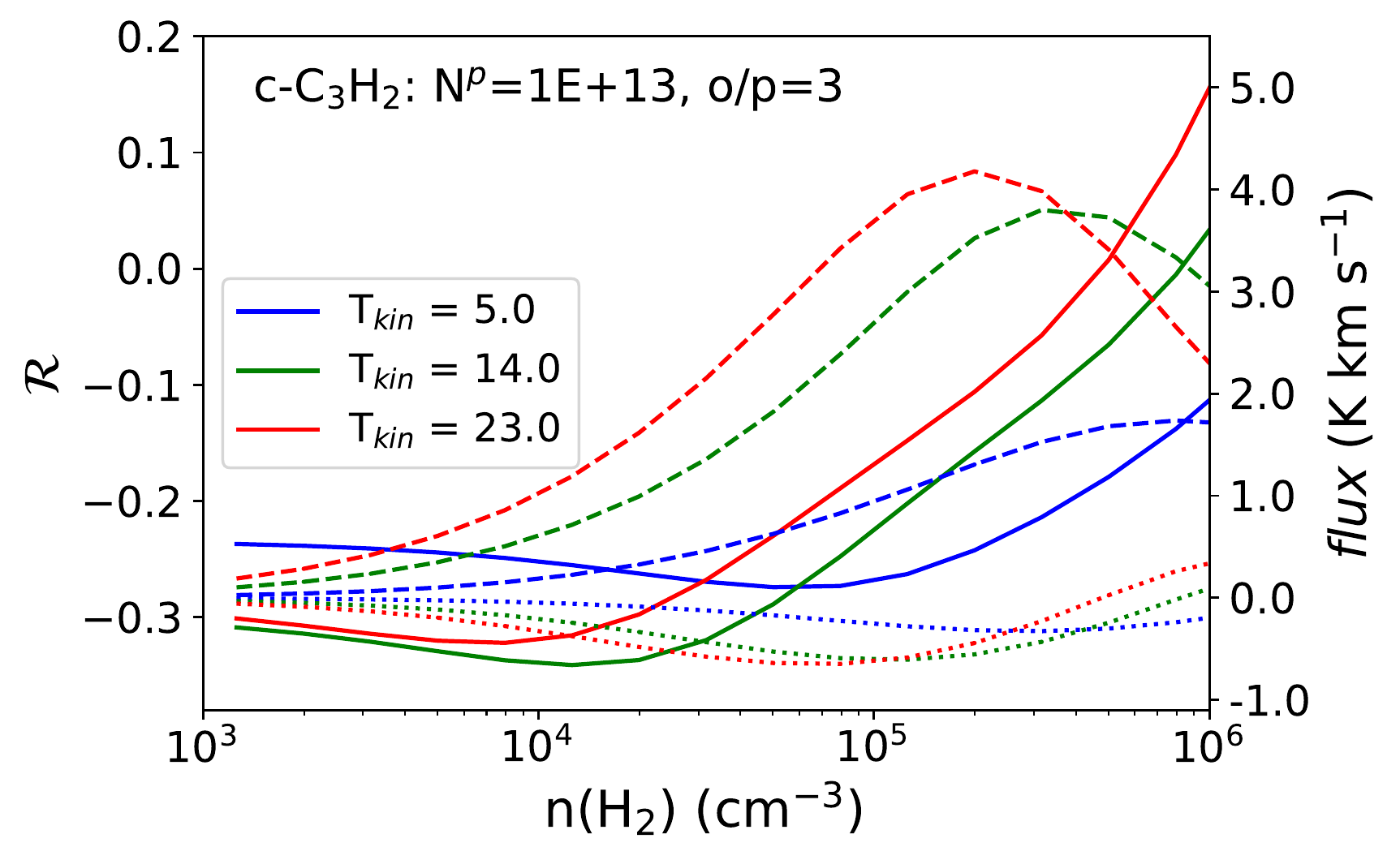}
\hspace{0.035\linewidth}
\caption{ Upper panel: The dependence between $\mathcal{R}$ 
(flux ratio between c-C$_3$H$_2$ 2$_{2,0}$--2$_{1,1}$ and c-C$_3$H$_2$ 1$_{1,0}$--1$_{0,1}$) 
and $n$(H$_2$) with the column density of $para$-C$_3$H$_2$ ($N^p$) fixed as 10$^{12}$ cm$^{-2}$.
Lower panel: Same as the upper panel except with $N^p$ fixed as 10$^{13}$ cm$^{-2}$.
The filled lines show the $\mathcal{R}$ values. The dashed and dotted lines show the fluxes of c-C$_3$H$_2$ 1$_{1,0}$--1$_{0,1}$
and c-C$_3$H$_2$ 2$_{2,0}$--2$_{1,1}$, respectively.
\label{radex_c3h2}}
\end{figure}

The two panels of Fig. \ref{radex_c3h2} show the relations between $\mathcal{R}$ and $n$ with 
$T_{\rm kin}$ of 5 K, 14 K and 23 K, 
and $N^p$ at $10^{11}$ cm$^{-2}$ (low column density case, hereafter LN) and 
$10^{13}$ cm$^{-2}$ (moderate column density case, hereafter MN).
In the LN case all transitions are optically thin and $\mathcal{R}$  increases monotonically  
with densities $n$(H$_2$).
$\mathcal{R}$ is always larger than $-0.45$ in this case.
In the MN case, the profiles of $\mathcal{R}$ along $n$(H$_2$) are similar to those in the LN case when 
$n$(H$_2$) $>10^5$ cm$^{-3}$.
However, when the density is low ($n$(H$_2$) $\sim10^3$ cm$^{-3}$), 
the absorption line 2$_{2,0}$--2$_{1,1}$ will be weak 
because  the molecules are accumulated in lower energy states and
the photons released from $2_{2,0}-1_{1,1}$ transition tend to be re-absorbed.
For the low density ($n$(H$_2$) $\sim10^3$ cm$^{-3}$) 
and moderate column density ($N^p=10^{13}$ cm$^{-2}$) case, the $\mathcal{R}$ 
 values are independent of kinetic
temperature and always larger than $-0.35$. 
Compared with the LN case, another characteristic of  the MN case   is that
$\mathcal{R}$  will first slightly decrease as the density $n$(H$_2$) increases.

In the LN case with kinetic temperature fixed at 10 K, the fitting between $\mathcal{R}$
and $n({\rm H}_2)$ gives
\begin{equation} 
\log(n({\rm H}_2))=\frac{ \ln(200\mathcal{R}+80)}{1.45}+3 \label{eq_R_rough}
\end{equation}
For $\mathcal{R}$ smaller than $-0.35$, 
the volume density should be smaller than or comparable with 10$^{4}$ cm$^{-3}$,
but can not be well constrained. For $\mathcal{R}$ larger than $-0.35$, 
the volume density should be larger than 10$^4$ cm$^{-3}$.
However, the assumption of the LN case may be not valid when the volume density is high, 
and Eq. \ref{eq_R_rough} can only serve as
a rough estimation of $n$(H$_2$).  

If the kinetic temperature is fixed and $\mathcal{R}$ is not too low (larger than $-0.35$), 
the free parameters $n$(H$_2$) and $N^p$, 
as well as the optical depths can be directly fitted from Eq. \ref{eq_R}.

\begin{figure}
\centering
\includegraphics[width=0.8\linewidth]{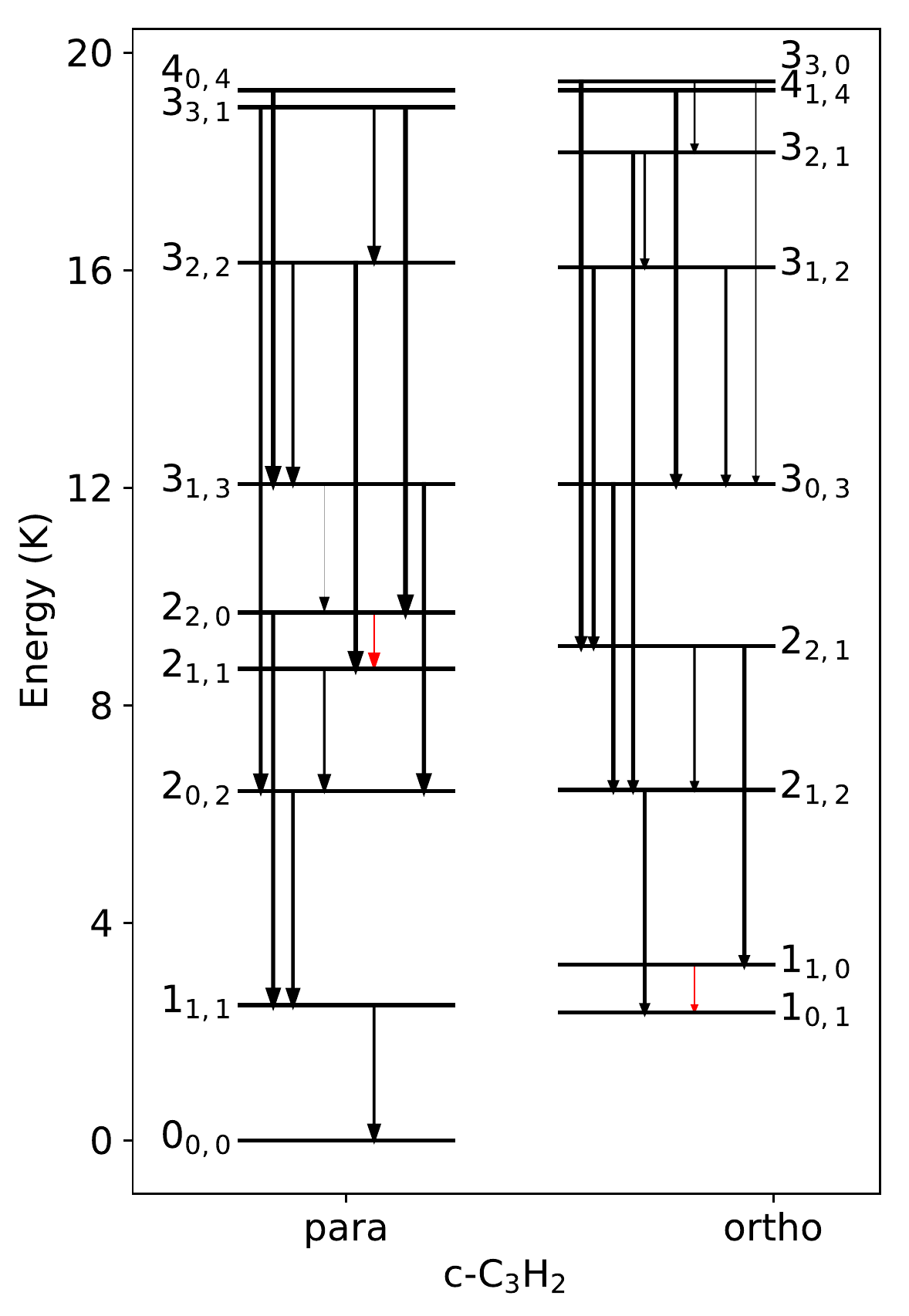}
\caption{The transitions of c-C$_3$H$_2$. The widths of the arrows
represent the 
corresponding Einstein  coefficients for spontaneous transition.
Red arrows: Observed transitions.
\label{c3h2_diag}
}
\end{figure}

\clearpage
\end{CJK}
\end{document}